\documentclass[referee,useAMS,usenatbib,usegraphicx]{mn2e}
\usepackage{color}

\newcommand{\hii}{\mbox{H~{\sc ii}~}}
\newcommand{\sii}{\mbox{[S~{\sc ii}]}~}
\newcommand{\hei}{\mbox{He~{\sc i}}}
\newcommand{\heii}{\mbox{He~{\sc ii}}}
\newcommand{\ciii}{\mbox{C~{\sc iii}}}
\newcommand{\siiv}{\mbox{Si~{\sc iv}}~}
\newcommand{\siii}{\mbox{Si~{\sc ii}}}
\newcommand{\siiii}{\mbox{Si~{\sc iii}}}
\newcommand{\oii}{\mbox{O~{\sc ii}}}
\newcommand{\mgii}{\mbox{Mg~{\sc ii}}~}
\newcommand{\nai}{\mbox{Na~{\sc i}}}
\newcommand{\cai}{\mbox{Ca~{\sc i}}}
\newcommand{\feii}{\mbox{Fe~{\sc ii}}}

\title[Stellar contents of Sh2-252 complex]{Optical and Near-infrared  survey of the stellar contents associated with the star-forming Complex Sh2-252}
\author[Jose et al.]{Jessy Jose $^{1,2}$\thanks{E-mail: jessy@iiap.res.in}, A.K. Pandey$^{1}$, K. Ogura$^3$, M.R. Samal$^{1,4}$, D.K. Ojha$^5$, B.C. Bhatt$^6$,
\newauthor
N. Chauhan$^{1,7}$, C. Eswaraiah$^{1}$, H. Mito$^{8}$, N. Kobayashi$^{8}$  and  R.K. Yadav$^1$\\
$^1$ Aryabhatta Research Institute of observational sciencES (ARIES), Manora Peak, Naini Tal, 263129, India\\
$^2$ Indian Institute of Astrophysics, Koramangala, Bangalore, 560 034, India\\
$^3$ Kokugakuin  University, Higashi, Shibuya-ku, Tokyo, 150-8440, Japan\\
$^4$ Laboratoire d'Astrophysique de Marseille - LAM, Universit\'e d'Aix-Marseille \& CNRS, UMR7326, 13388 Marseille Cedex 13, France\\
$^5$ Tata Institute of Fundamental Research, Mumbai (Bombay), 400 005, India\\
$^6$ CREST, Indian Institute of Astrophysics, Koramangala, Bangalore, 560 034, India\\
$^7$ Institute of Astronomy, National Central University, Chung-Li 32054, Taiwan\\
$^8$ Kiso Observatory, School of Science, University of Tokyo, Mitake, Kiso-machi, Kiso-gun, Nagano-ken 397-0101, Japan\\
}
\begin{document}

\date{}

\pubyear{2012}

\maketitle

\label{firstpage}

\begin{abstract}

We present the analyses  of the stellar contents associated with the extended \hii region Sh2-252 using  deep optical $UBVRI$ photometry, 
slit and slitless spectroscopy along with the near-infrared (NIR) data from 2MASS for an area $\sim$ 1 degree $\times$ 1 degree.
We have studied the sub-regions of Sh2-252 which includes four compact-\hii  (C\hii) regions, namely A, B, C and E and 
two clusters NGC 2175s and Teutsch 136 (Teu 136). Of the fifteen  spectroscopically observed bright stars, eight have been 
identified as massive members of spectral class earlier than B3. From the spectro-photometric analyses, we derived the average 
distance of the region as 2.4$\pm$0.2 kpc and the reddening $E(B-V)$ of the massive members is found to vary between 0.35 to 2.1 mag. 
We found  that  NGC 2175s and Teu 136, located towards the eastern edge of the complex  are the sub-clusters 
of Sh2-252.  The stellar surface density distribution in $K$-band shows  clustering associated with the regions A, C, E, NGC 2175s and 
Teu 136. We have also identified the candidate ionizing sources of the C\hii regions.  61 H$\alpha$ emission sources are identified 
using slitless spectroscopy. The distribution of the  H$\alpha$ emission sources and  candidate 
young stellar objects (YSOs) with IR excess on the $V/(V-I)$ colour magnitude diagram (CMD) shows that a majority of them have 
approximate  ages between 0.1 - 5 Myr and  masses in the range of  0.3 - 2.5 M$_{\odot}$.  The optical CMDs of the  candidate 
pre-main sequence (PMS) sources in the  individual regions also show an age spread  of  0.1 - 5 Myr for each of them. 
We  calculated  the $K$-band luminosity functions (KLFs) for the sub-regions  A, C, E, NGC 2175s and Teu 136. Within errors, 
the KLFs for all the sub-regions are found to be similar and  comparable to that of young clusters 
of age $<$ 5 Myr.  We also estimated  the mass functions (MFs) of the PMS sample of the individual regions in  the 
mass range of $0.3 - 2.5 M_{\odot}$. In general, the slopes of the MFs of all the sub-regions  are found comparable to the Salpeter value.

\end{abstract}

\begin{keywords}
stars: formation $-$ stars: luminosity function, mass function $-$ stars:
pre$-$main$-$sequence $-$ \hii regions: individual: Sh2-252
\end{keywords}

\section{Introduction}

It is known that most of the stars in the galaxy  are formed in clusters or in OB associations where massive members
of the region significantly affect their environment. This is possible in the form of
strong stellar winds, intense UV radiation and eventually supernova explosions. These processes
can ultimately cause their natal molecular cloud to be destroyed, which, in turn, will put
an end to further star formation. However, it has also been observed that the above processes
can trigger the birth of new generations of stars (e.g., Elmegreen \& Lada 1977) and is likely
to influence the key properties  of the star formation process such as the initial mass function
(IMF), star formation efficiency and the evolution of protostellar disks around the young stars (e.g., Clarke 2007). 
Keeping in mind the above points, it is important to study the large scale properties of OB associations so that
we can have better understanding of the various processes governing the star formation.

As a continuation of our multi-wavelength analyses of star forming regions (SFRs; e.g.,
Pandey et al. 2008, Jose et al. 2008; 2011), an optically bright \hii region Sh2-252 
(Sharpless 1959; $\alpha_{2000}$ = $06^{h}09^{m}39^{s}$; $\delta_{2000}$ = $+20^{\circ}29^{\prime}12^{\prime\prime}$;  
l=190.04;  b=+0.48) has been studied  in this paper. It is located in the galactic anti-center direction  
and is a part of the Gemini OB1 association. Radio and CO surveys have  already been
done for this region by many authors (Felli et al. 1977, Lada \& Wooden 1979,
K\"{o}mpe et al. 1989; Szymczak et al. 2000 etc.). However, stellar contents of this
region was poorly studied in optical and infrared wavebands.  Also, authors have raised arguments regarding the presence of many
ionizing sources in this complex  (see Felli et al. 1977, Lada \& Wooden 1979).
However, the massive members of this region have not yet been characterized spectroscopically except 
three bright sources. In this paper, we use the deep optical data in $UBVRI$
bands, slit and slitless spectroscopic observations along with the $JHK$ data from
2MASS to identify and classify the massive members, to determine the reddening, 
distance  of the region as well as the age and mass of candidate pre-main sequence
(PMS) sources and finally  to obtain the K-band luminosity function (KLF) and
IMF. After a brief description of the previous studies on Sh2-252, section 2 describes various data sets
used for the present study. Analyses and results including fundamental parameters of Sh2-252, 
identification and classification of massive stars  in the associated clusters, its various properties
such as age, distance, reddening, KLF and  IMF  are  discussed in section 3 and section
4 summarizes the results.

\subsection{Overview of previous studies on Sh2-252}
\label{overview}

\begin{figure*}
\centering
\includegraphics[scale = 0.86, trim =0 20 20 230, clip]{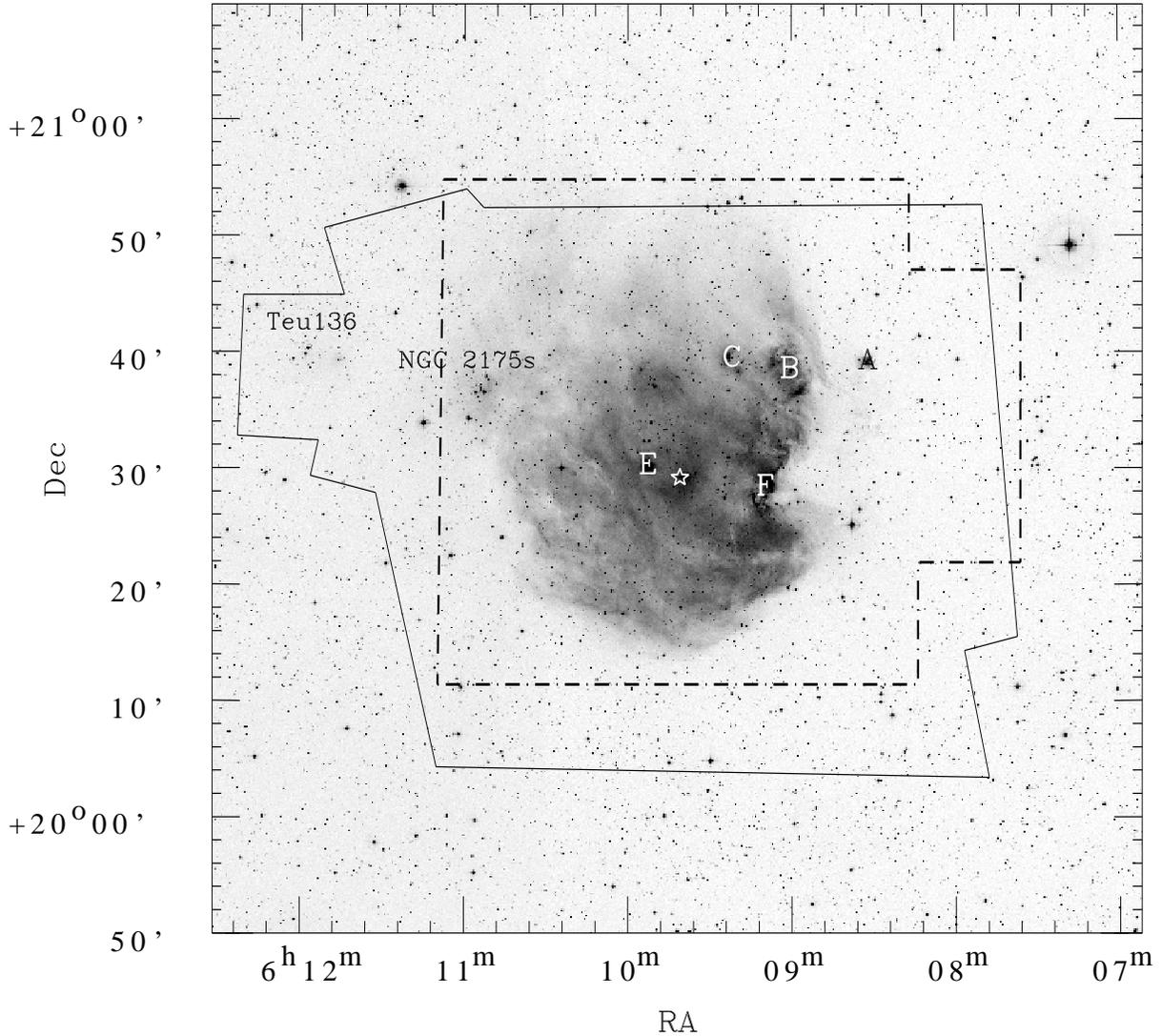}
\vspace{2mm}
\caption{The DDS2-R band image of the region around Sh2-252  for an area of 
$\sim 1.3  \times 1.3$  square degrees. The locations of the thermal radio sources
of the region A, B, C, E and F identified by Felli et al. (1977) along with  the small clusters NGC 2175s and  
Teu 136 are marked in the figure. The ionizing source (HD 42088) of the \hii region is marked using a 
star symbol. The area covered for the deep optical observations in $V$ and $I$ bands is shown using 
continuous lines (polygon) and the thick dot-dashed  box represents the area covered with the 
$H{\alpha}$ slitless spectroscopy survey. The abscissa and the ordinates are for the J2000 epoch. }
\label{area}
\end{figure*}

\begin{figure*} 
\centering
\includegraphics[scale = 0.93, trim = 0 0 0 0, clip]{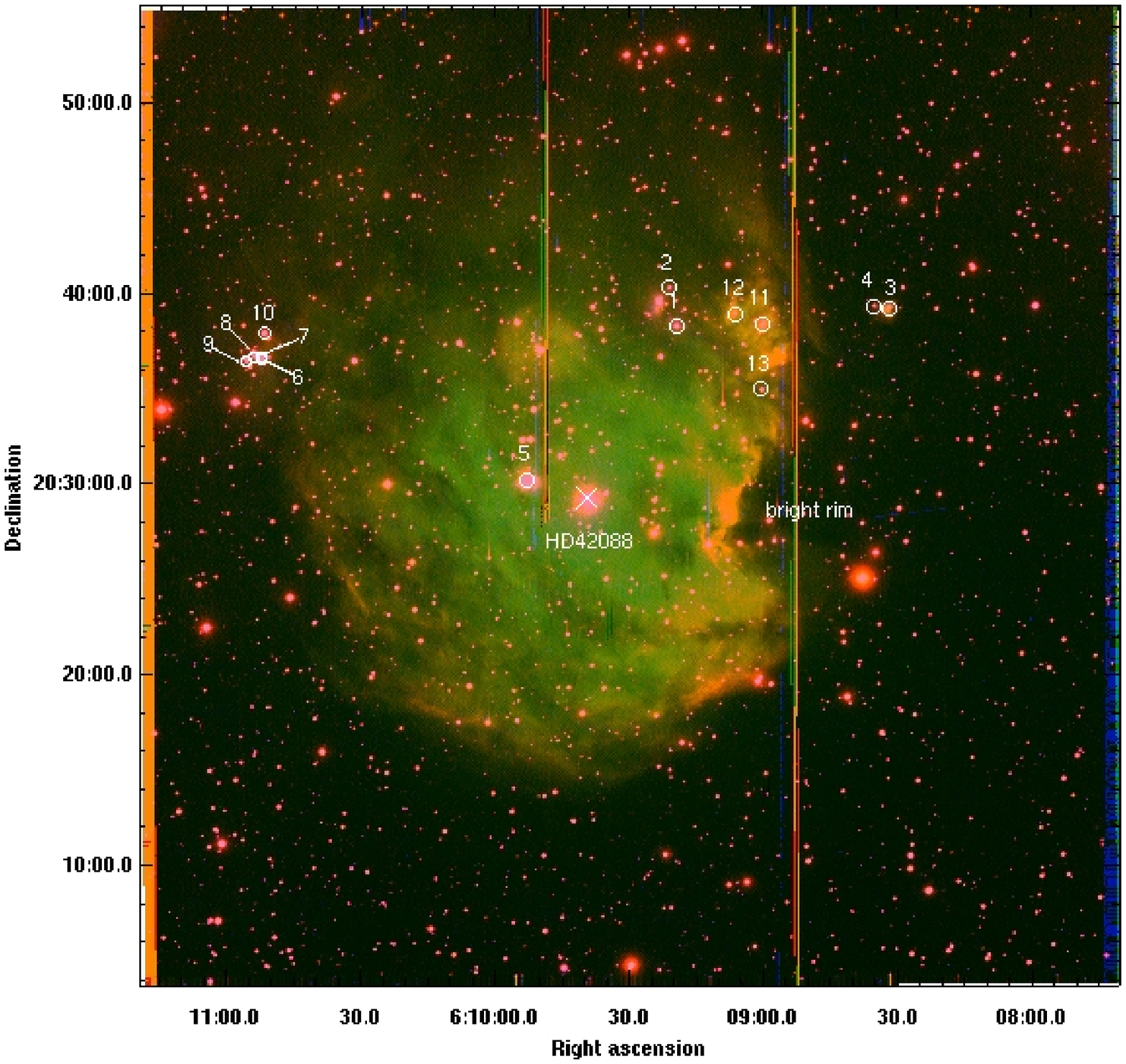}

\caption {A close-up colour composite view of Sh2-252 reproduced by using the  $V$, $H{\alpha}$  and
\sii bands ($V$: blue; $H{\alpha}$: green;  \sii: red)  for an area  $\sim 50 \times 50$ arc min$^2$ 
obtained from the Kiso observations. The sources numbered  are further classified by using spectroscopic
observations (see section \ref{spectra}) and their identification are also shown in the figure. The 
locations of the ionizing source HD 42088 and the 
bright rim are also marked.  The lines in the image are artifacts.  }

\label{color-opt}
\end{figure*}

The low density (n$_e$ = 9 $cm^{-3}$)  and the  relatively large  size (D $\sim$ 20 pc)  of the extended region place Sh2-252 in 
the class of evolved \hii regions (Churchwell 1974). 
 Main source of ionization of Sh2-252 is the central star HD 42088 with a spectral type O6.5 V (Conti \& Alschular 1971; 
Walborn 1972), which is a member of the Gemini OB1 association. In Fig. \ref{area}, the 
DSS2-R band image of the region around Sh2-252 for an area of $\sim 1.3  \times 1.3$  square degrees is given, 
which shows a bright nebula having a diameter $\sim$ 30$^\prime$  with a bright rim (Felli et al. 1977) at the west. See the colour
composite image given in Fig. \ref{color-opt} for a close up view. Two stellar clusters  were identified in this 
region optically, one is  the dispersed cluster (namely NGC 2175) centered on the nebulosity  and the other 
of smaller diameter (namely NGC 2175s) is
located at the north-east of the nebula. Pismis (1970) estimated the distances to these clusters as 1900 pc and 3500 pc, respectively
and hence NGC 2175s was thought  to be a background object.  Subsequently, using eight colour photometry, 
Chavarr\'{i}a et al. (1987)  found that these two  clusters are located at a distance of 2.0 kpc and hence claimed that both are 
associated with Sh2-252. A new near-infrared (NIR) cluster Teutsch 136 (here after Teu 136; see Fig. \ref{area}) has been identified towards the 
east of NGC 2175s by Koposov et al. (2008). However, they did not make any parameter determination for this
cluster. Hence, the association of this object with Sh2-252
needs to be checked. In the literature, the distance to Sh2-252 from spectrophotometric and kinematic studies vary 
between 1.4 kpc to 2.6 kpc (e.g., Bonatto \& Bica 2011,
Pismis 1970, Reid et al. 2009 and  Grasdalen \& Carrasco 1975).  Using the 5 GHz aperture synthesis observations,
Felli et al. (1977) detected six extended radio sources towards Sh2-252 labeled as Sh2-252 A to F 
(see Fig. 1 in Lada \& Wooden 1979). They have identified four of these sources (A, B, C and E) as 
compact \hii (C\hii) regions, probably having  a local source of ionization in each of them with a spectral type later 
than HD 42088 (i.e., O6.5V). The region F  coincides with the optical bright rim (see Fig. \ref{area}, 
and see also Fig. \ref{color-opt} for a close up view) and no internal  heating source is found within 
it (Lada \& Wooden 1979).  The molecular survey by Lada \& Wooden (1979) shows that the  most intense CO peak 
is located very close to  Sh2-252A with  water and methanol maser emissions  within its  proximity (Lada \& Wooden 1979; 
K\"{o}mpe et al. 1989; Szymczak et al. 2000), which  indicates  the recent  star formation
activity towards this region.  Most of the region around Sh2-252A seems to be optically obscured, however, 
there is an embedded cluster visible in NIR and recently, Tej et al. (2006) has studied this cluster in detail.

\section{OBSERVATIONS AND DATA REDUCTIONS}

\subsection{Optical CCD imaging}
\label{obs}
The CCD $UBVRI$ observations of Sh2-252 were carried out using the 105-cm Schmidt Telescope of the Kiso
observatory, Japan, on 2004 November 4. The 2K $\times$ 2K CCD with each pixel corresponding to 1.5 arcsec 
covers a field of $\sim 50\times50$ arcmin$^2$ in the sky. The log of observations is tabulated in 
Table \ref{obslog}.


\begin{table}
\caption{Log of observations}
\label{obslog}
\scriptsize
\begin{center}
\begin{tabular}{|p{.60in}p{.7in}p{.78in}p{2.6in}|}
\hline
$\alpha_{(2000)}$ & $\delta_{(2000)}$ & Date of & Filter \& Exp. time (sec) $\times$ no. of frames\\
(h:m:s) & (d:m:s)  &    observation  &             \\
\hline
  {\it Kiso$^1$}                 &           &   &\\

06:09:32& +20:28:10& 2004.11.04 & $U$:  60$\times$2; B: 20$\times$2; V: 10$\times$2; R: 10$\times$2; I: 10$\times$2  \\

{\it ST$^2$} &                  &            &\\
06:09:39& +20:29:12&  2005.12.02 & $U$:  360$\times$2; B: 240$\times$2; V: 120$\times$2; R: 60$\times$2; I: 60$\times$2  \\
06:10:09& +20:34:57&  2006.10.19 & $V$:  600$\times$6; $I$: 300$\times$6\\
06:09:14& +20:35:60&  2006.10.23 & $V$:  600$\times$6; $I$: 300$\times$6\\
06:10:09& +20:25:28&  2006.10.23 & $V$:  600$\times$6; $I$: 300$\times$6\\
06:09:18& +20:24:52&  2006.12.06 & $V$:  600$\times$6; $I$: 300$\times$6\\
06:08:06& +20:20:20&  2007.01.15 & $V$:  600$\times$6; $I$: 300$\times$6\\
06:08:06& +20:20:20&  2007.01.15 & $V$:  600$\times$6; $I$: 300$\times$6\\
06:10:56& +20:35:29&  2007.01.15 & $V$:  600$\times$6; $I$: 300$\times$6\\
06:09:11& +20:25:22&  2007.11.06 & $V$:  600$\times$6; $I$: 300$\times$6\\
06:10:54& +20:36:18&  2007.11.06 & $V$:  600$\times$6; $I$: 300$\times$6\\
06:08:51& +20:12:36&  2008.01.08 & $V$:  600$\times$6; $I$: 300$\times$6\\
06:11:56& +20:38:28&  2008.01.08 & $V$:  600$\times$6; $I$: 300$\times$6\\
06:10:25& +20:13:46&  2008.04.15 & $V$:  600$\times$6; $I$: 300$\times$6\\
06:11:31& +20:34:35&  2008.04.15 & $V$:  600$\times$6; $I$: 300$\times$6\\
06:08:31& +20:39:24&  2008.12.18 & $V$:  600$\times$6; $I$: 300$\times$6\\
06:09:29& +20:46:24&  2008.12.18 & $V$:  600$\times$6; $I$: 300$\times$6\\
06:10:24& +20:45:31&  2008.12.18 & $V$:  600$\times$6; $I$: 300$\times$6\\
06:09:42& +20:13:32&  2008.12.18 & $V$:  600$\times$6; $I$: 300$\times$6\\
06:11:09& +20:24:36&  2008.12.19 & $V$:  600$\times$6; $I$: 300$\times$6\\
06:08:23& +20:26:44&  2008.12.19 & $V$:  600$\times$6; $I$: 300$\times$6\\
06:11:16& +20:46:54&  2008.12.19 & $V$:  600$\times$6; $I$: 300$\times$6\\

{\it HCT$^3$} &                  &             &    \\
06:09:21& +20:40:37&  2005.01.07 & Gr5/H$\alpha$-Br:  450$\times$3; H$\alpha$-Br:   90$\times$2\\
06:09:21& +20:24:31&  2005.01.08 & Gr5/H$\alpha$-Br:  450$\times$3; H$\alpha$-Br:   90$\times$2\\
06:09:60& +20:32:38&  2005.01.09 & Gr5/H$\alpha$-Br:  450$\times$3; H$\alpha$-Br:   90$\times$2\\
06:09:60& +20:40:39&  2005.01.09 & Gr5/H$\alpha$-Br:  450$\times$3; H$\alpha$-Br:   90$\times$2\\
06:10:37& +20:40:46&  2005.01.09 & Gr5/H$\alpha$-Br:  600$\times$3; H$\alpha$-Br:   120$\times$2\\
06:10:37& +20:32:55&  2005.02.16 & Gr5/H$\alpha$-Br:  600$\times$3; H$\alpha$-Br:   120$\times$2\\
06:10:37& +20:24:28&  2005.02.16 & Gr5/H$\alpha$-Br:  600$\times$3; H$\alpha$-Br:   120$\times$2\\
06:09:23& +20:48:58&  2005.02.17 & Gr5/H$\alpha$-Br:  600$\times$3; H$\alpha$-Br:   120$\times$2\\
06:09:23& +20:56:58&  2005.02.17 & Gr5/H$\alpha$-Br:  600$\times$3; H$\alpha$-Br:   120$\times$2\\
06:10:24& +20:15:44&  2005.02.17 & Gr5/H$\alpha$-Br:  600$\times$3; H$\alpha$-Br:   120$\times$2\\
06:09:23& +20:16:52&  2005.02.17 & Gr5/H$\alpha$-Br:  600$\times$3; H$\alpha$-Br:   120$\times$2\\
06:10:54& +20:36:23&  2006.01.09 & Gr5/H$\alpha$-Br:  600$\times$3; H$\alpha$-Br:   150$\times$2\\
06:08:43& +20:40:47&  2007.01.25 & Gr5/H$\alpha$-Br:  600$\times$3; H$\alpha$-Br:   120$\times$2\\
06:08:43& +20:32:52&  2007.01.25 & Gr5/H$\alpha$-Br:  600$\times$3; H$\alpha$-Br:   120$\times$2\\
06:08:43& +20:24:53&  2007.01.25 & Gr5/H$\alpha$-Br:  600$\times$3; H$\alpha$-Br:   120$\times$2\\
06:08:05& +20:32:57&  2007.01.25 & Gr5/H$\alpha$-Br:  450$\times$3; H$\alpha$-Br:   90$\times$2\\
06:08:05& +20:40:56&  2007.01.25 & Gr5/H$\alpha$-Br:  450$\times$3; H$\alpha$-Br:   90$\times$2\\
06:08:43& +20:16:58&  2007.01.26 & Gr5/H$\alpha$-Br:  450$\times$3; H$\alpha$-Br:   90$\times$2\\
06:08:05& +20:25:01&  2007.01.26 & Gr5/H$\alpha$-Br:  450$\times$3; H$\alpha$-Br:   90$\times$2\\
06:08:05& +20:33:01&  2007.01.26 & Gr5/H$\alpha$-Br:  450$\times$3; H$\alpha$-Br:   90$\times$2\\
06:09:46& +20:38:44&  2009.11.15 & Gr5/H$\alpha$-Br:  600$\times$3; H$\alpha$-Br:   90$\times$2\\
06:09:05& +20:38:44&  2009.11.15 & Gr5/H$\alpha$-Br:  600$\times$3; H$\alpha$-Br:   90$\times$2\\
06:09:46& +20:28:53&  2009.11.15 & Gr5/H$\alpha$-Br:  600$\times$3; H$\alpha$-Br:   90$\times$2\\
06:08:31& +20+53:16&  2009.11.15 & Gr5/H$\alpha$-Br:  600$\times$3; H$\alpha$-Br:   90$\times$2\\
06:10:14& +20+52:54&  2009.11.15 & Gr5/H$\alpha$-Br:  600$\times$3; H$\alpha$-Br:   90$\times$2\\

\hline
\end{tabular}
\end{center}
$^1$  105-cm Schmidt Telescope, Kiso, Japan\\
$^2$  104-cm Sampurnanand Telescope, ARIES, Naini Tal\\
$^3$  2-m Himalayan Chandra Telescope, IAO, Hanle\\

\end{table}


The CCD $UBVRI$ photometry of the central region  ($\alpha_{2000}$ = $06^{h}09^{m}39^{s}$; $\delta_{2000}$ = 
$+20^{\circ}29^{\prime}12^{\prime\prime}$) of Sh2-252 was carried out using the 104-cm Sampurnand Telescope 
(ST) of ARIES, Nainital, India on 2005 December 2. The 
2K $\times$ 2K CCD with a plate scale of 0.37 arcsec pixel$^{-1}$  covers a field of $\sim 13\times13$ arcmin$^2$ 
in the sky.  To improve the  signal to noise  ratio, the observations were carried out in a binning mode of
$2\times2$ pixels. The standard field SA 98 from Landolt (1992) was observed on the same night to  apply the atmospheric 
and instrument corrections to the target field.  We then performed PSF photometry with the DAOPHOT package in IRAF, 
and derived the  extinction coefficients and colour coefficients  using the observations of standard field SA 98. 
The central region 
of Sh2-252 observed with ST was calibrated by applying these coefficients. The calibration uncertainties 
between the standard and transformed  $V$ magnitudes and  $U-B$, $B-V$, $V-R$, $V-I$ colours  were $\le$
0.03 mag. The selected isolated stars of this central calibrated region  were used  to calibrate  the wide 
field observations taken with  Kiso Schmidt Telescope. 

We  further carried out  deep  observations in $V$ and $I$ bands using ST for 20 overlapping 
sub-regions of Sh2-252, to cover an area $\sim$  1 degree $\times$ 1 degree around it. The observations 
were conducted from  2006 October to 2008 December in 
good photometric conditions and the log of observations is tabulated in Table \ref{obslog}. Each region was exposed 
for 60 min in $V$ and 30 min in $I$-band. The combined area of these 20 overlapping regions is shown in Fig. \ref{area}
using   solid lines.  The secondary standards 
from the Kiso observations were used to calibrate these individual regions. The final catalog is made from these 
three sets of observations, where, the magnitudes of the bright ($V<$ 15 mag) and  faint ($V>$ 15 mag)  
objects were taken from Kiso observations 
and  ST observations, respectively. To ensure good photometric quality, we selected
only those sources having uncertainty $<$0.2 mag in $V$-band to make the final catalog. Thus  we obtained 
photometry of 8791  sources detected at least  in $V$ and $I$-bands with a limiting magnitude of $V$ $\sim$23 mag. 

In order to check the photometry accuracy, we compared our photometry with the $UBV$ photometries available in the
literature  by  Grasdalen \& Carrasco (1975)  and Haikala (1994). The mean  difference between our 
photometry and  theirs were  $\le$ 0.03 mag in $V, B-V$ and $U-B$, which shows that our photometry 
is in agreement with the previous studies.

\subsection{$H{\alpha}$ slitless grism spectroscopy}
\label{slitless}

An $H{\alpha}$ emission line survey of Sh2-252 was conducted between 2005 January and 2009 November. 
The observations were done  in the slitless mode with a grism as the dispersing element using the HFOSC
of  Himalayan Chandra Telescope (HCT) of IAO, Hanle, India. The central 2K $\times$ 2K  pixels of the  2K $\times$ 4K CCD 
were used for data acquisition. The CCD  with an image scale of 0.296 arcsec pixel$^{-1}$  covers an 
area of $\sim$ 10 $\times$  10 arcmin$^2$ in the sky. A combination of $H{\alpha}$ broad-band  filter 
($H{\alpha}$-Br; 6100 - 6740 {\AA}) and Grism 5 were used. The resolution of Grism 5 
is 870. We observed 25 overlapping sub-regions around Sh2-252 and the area  covered 
for this observations is shown in Fig. \ref{area} using a dot-dashed box. Multiple frames 
were taken to ensure the presence  of $H{\alpha}$ emitting sources. The log of the observations 
are given in Table \ref{obslog}. $H{\alpha}$ emission line stars with an enhancement over the continuum at 
the $H{\alpha}$ line were visually identified. We obtained 61 $H{\alpha}$ emitting sources within our surveyed area.  
The detection limit of  the H-alpha survey  is about 3 {\AA} in terms of equivalent width, or V $\sim$ 22 in terms of magnitude.
In Table \ref{hadata}, we have provided the J2000 coordinates, $V$-band magnitudes and $V-I$ colours
for these emission line sources, if available.


\begin{table*}

\caption{photometric data of $H{\alpha}$ emission line sources in Sh2-252}
\label{hadata}
\begin{center}
\scriptsize
\begin{tabular}{|p{.80in}p{.85in}p{.85in}p{.5in}p{.5in}p{.5in}p{.5in}|}

 \hline

 $\alpha_{(2000)}$& $\delta_{(2000)}$& $V$ &$V_{err}$ &$(V-I)$ & $(V-I)_{err}$ \\
 (h:m:s)        &  ($^{\circ}:^{\prime}:^{\prime\prime}$)   &     &        &        &    \\
\hline
	  06:07:48.61& +20:39:16.3  &	-	&   -	   &  -      &  -          \\
	  06:07:59.15& +20:29:19.3  &  21.187  &   0.053  &   2.970 &    0.054    \\
	  06:08:06.25& +20:33:47.0  &  18.927  &   0.010  &   1.487 &    0.015    \\
    06:08:30.31& +20:37:18.7  &   -	&   -	   &  -      &  -          \\
	  06:08:31.77& +20:31:42.2  &  19.754  &   0.040  &   3.147 &    0.051    \\
	  06:08:33.26& +20:40:38.6  &  18.049  &   0.028  &   2.503 &    0.056    \\
    06:08:35.19& +20:30:22.7  &   -	&   -	   &  -      &  -          \\
    06:08:46.41& +20:39:00.4  &   -	&   -	   &  -      &  -          \\
    06:08:47.70& +20:29:06.0  &   -	&   -	   &  -      &  -          \\
    06:08:50.44& +20:36:41.4  &  22.081 &    0.078 &    3.546&     0.080\\
	  06:08:51.26& +20:35:36.0  &  19.797  &   0.053  &   2.774 &    0.070 \\
	  06:08:51.59& +20:35:37.1  &  18.179  &   0.007  &   2.371 &    0.012 \\
	  06:08:52.62& +20:37:28.7  &  18.200  &   0.008  &   2.120 &    0.012 \\
	  06:08:55.20& +20:39:33.3  &  16.876  &   0.007  &   1.501 &    0.011 \\
	  06:08:57.21& +20:38:45.2  &  19.694  &   0.014  &   2.315 &    0.018 \\
	  06:09:01.05& +20:43:10.8  &  19.455  &   0.012  &   2.604 &    0.014 \\
	  06:09:02.77& +20:36:43.4  &  14.427  &   0.004  &   0.847 &    0.014 \\
	  06:09:13.22& +20:38:20.6  &  20.888  &   0.029  &   2.791 &    0.032 \\
    06:09:14.65& +20:41:26.9  &   -	&   -	   &  -      &  -      \\
	  06:09:21.23& +20:40:22.6  &  14.112  &   0.011  &   1.365 &    0.016\\
	  06:09:21.31& +20:38:10.0  &  20.339  &   0.026  &   2.960 &    0.030\\
	  06:09:22.16& +20:41:58.1  &  17.371  &   0.009  &   1.975 &    0.026\\
	  06:09:22.64& +20:36:10.2  &  19.579  &   0.023  &   2.549 &    0.033\\
	  06:09:23.43& +20:38:02.7  &  18.914  &   0.008  &   2.456 &    0.014\\
	  06:09:24.04& +20:36:44.8  &  20.415  &   0.079  &   3.276 &    0.094\\
	  06:09:25.36& +20:37:11.0  &  19.665  &   0.017  &   2.968 &    0.019\\
	  06:09:25.54& +20:36:36.0  &  19.817  &   0.019  &   2.605 &    0.021\\
	  06:09:26.31& +20:38:09.4  &  19.902  &   0.015  &   2.920 &    0.019 \\
	  06:09:26.77& +20:37:12.7  &  19.628  &   0.026  &   2.430&     0.042 \\
	  06:09:27.19& +20:30:32.6  &  17.900  &   0.004  &   1.990 &    0.008 \\
	  06:09:27.24& +20:37:51.6  &  15.865  &   0.007  &   1.484 &    0.011 \\
    06:09:28.57& +20:37:34.8 &   -	&   -	   &  -      &  -       \\
	  06:09:29.56& +20:41:26.7  &  20.462  &   0.022  &   2.532 &    0.034 \\
	  06:09:30.50& +20:37:46.4  &  20.855  &   0.031  &   2.834 &    0.034 \\
	  06:09:31.39& +20:38:31.8  &  19.913  &   0.014  &   2.279 &    0.024 \\
	  06:09:34.27& +20:44:11.6  &  18.810  &   0.007  &   2.805 &    0.010 \\
	  06:09:35.25& +20:31:15.3  &  19.371  &   0.013  &   2.075 &    0.021 \\
	  06:09:37.71& +20:37:55.8  &  20.013  &   0.014  &   2.731 &    0.019 \\
	  06:09:39.30& +20:38:43.6  &  21.565  &   0.060  &   2.546 &    0.069 \\
	  06:09:39.38& +20:34:28.9  &  19.130  &   0.010  &   2.231 &    0.013 \\
	  06:09:41.35& +20:43:37.7  &  17.666  &   0.006  &   1.729 &    0.010 \\
	  06:09:41.77& +20:31:25.0  &  19.140  &   0.013  &   2.421 &    0.018 \\
	  06:09:51.52& +20:28:33.1  &  20.623  &   0.046  &   3.078 &    0.049 \\
	  06:09:51.67& +20:30:06.6  &  14.373  &   0.017  &   1.066 &    0.025 \\
	  06:09:55.35& +20:33:35.0  &  17.166  &   0.004  &   2.254 &    0.006 \\
	  06:09:55.44& +20:40:13.7  &  15.980  &   0.003  &   1.613 &    0.005\\
	  06:10:05.45& +20:35:53.7  &  18.218  &   0.004  &   1.902 &    0.007\\
	  06:10:14.97& +20:43:22.5  &  15.356  &   0.005  &   1.340 &    0.007\\
	  06:10:35.48& +20:29:26.3  &  18.628  &   0.010  &   1.795 &    0.033\\
	  06:10:37.84& +20:33:42.4  &  19.289  &   0.007  &   2.267 &    0.009\\
	  06:10:38.58& +20:39:24.9  &  17.006  &   0.008  &   1.614 &    0.018\\
	  06:10:39.95& +20:33:36.4  &  19.517  &   0.009  &   2.196 &    0.011\\
	  06:10:43.60& +20:35:41.1  &  18.487  &   0.005  &   2.320 &    0.011\\
	  06:10:45.36& +20:41:55.3  &  19.274  &   0.007  &   2.494 &    0.008\\
	  06:10:46.37& +20:34:06.0  &  18.500  &   0.004  &   2.290 &    0.006 \\
	  06:10:48.22& +20:31:45.7  &  19.143  &   0.018  &   2.250 &    0.023 \\
	  06:10:49.02& +20:35:23.9  &  20.078  &   0.015  &   1.912 &    0.020 \\
	  06:10:49.94& +20:36:59.0  &  18.485  &   0.005  &   1.968 &    0.007 \\
	  06:10:50.96& +20:31:05.8  &  21.582  &   0.121  &   3.213 &    0.123 \\
	  06:10:53.67& +20:32:23.6  &  13.556  &   0.005  &   1.104 &    0.005 \\
 	  06:10:57.60& +20:42:46.3  &  18.427  &   0.006  &   1.866 &    0.010 \\
 
\hline
\end{tabular}\\
\end{center}
\end{table*}

\subsection{Slit  spectroscopy}
\label{slit}

Low resolution optical spectroscopic observations of  fifteen optically bright sources  in Sh2-252 were made
using the HFOSC of HCT. These sources are the probable massive members of Sh2-252 and some of them are the
candidate ionizing sources of the sub-\hii regions (see section \ref{isource}). The locations of these 
sources are marked in Fig. \ref{color-opt}.  The observations were taken on 16 and 17 November 2009.  
The  spectra in the wavelength range  3800-6840 {\AA} with a dispersion of 1.45  {\AA}  pixel$^{-1}$ were obtained
using the  low  resolution Grism 7 with a slit of  2 arcsec width  and exposure time of 900s. One-dimensional spectra
were  extracted from the  bias-subtracted and flat-field corrected images by using the optimal
extraction method in IRAF. Wavelength calibration of the spectra was done by using an FeAr 
lamp source. Spectro photometric standard star (G191B2B)  was observed on these two nights to give the flux
calibration  to the target spectra.

\subsection{Near-infrared data from 2MASS}

NIR $JHK$ data for  point sources within a radius of 30 arcmin
around Sh2-252  have been obtained from  2MASS point source catalog (PSC) (Cutri et al. 2003).  To improve
photometric accuracy, we used photometric quality flag (ph$\_$qual =
AAA) which gives a S/N $\ge$ 10 and a  photometric uncertainty $ <$
0.10 mag. This selection criterion ensures best quality detection in
terms of photometry and astrometry (cf. Lee et al. 2005).
The $JHK$ data  were transformed  from the 2MASS
system  to the California Institute of Technology (CIT)  system by using
the relations given  by Carpenter (2001).

\subsection{Completeness of the data}
\label{cf}

To study  the luminosity function (LF) / mass function (MF), it  is  necessary to  take into  account  the incompleteness
of  the  data that could  occur due to various factors (e.g., crowding of the stars). We used
ADDSTAR  routine of DAOPHOT to determine the completeness factor (CF). The procedure has been
outlined in  detail in our earlier works (see e.g., Pandey et al. 2001). Briefly, we randomly added
artificial stars to both $V$ and $I$ images taken with ST in such a way that they have similar
geometrical locations but differ in $I$ brightness according to the mean $(V-I)$ colour ($\sim 1.5$ mag)
of the data sample. The luminosity distribution of artificial stars was chosen in such a way that more
number of stars were inserted towards the fainter magnitude bins. The frames were reduced using the
same procedure used for the original frames. The ratio of the number of stars recovered to those added
in each magnitude interval gives the CF as a function of magnitude. The minimum value of the CF of the
pair (i.e., $V$- and $I$-bands) was used to correct the data incompleteness.  It is to be noted that
the spatially varying nebular background (see. Fig. \ref{color-opt}) and stellar crowding characteristics (see Fig. \ref{cont})
could affect the local completeness limit. In order to account for this, we estimated the CFs of two representative sub-regions
Sh2-252C and NGC 2175s and the values obtained are given in Table \ref{cftable}. The CFs of both the regions seem to be 
agreeing well with each other except for the faint magnitude bin, where the  variation seems to be significant. 

We also estimated the CFs of the 2MASS data using the $K$-band images from archive around the regions Sh2-252C and NGC 2175s. 
Using the DAOPHOT $/$ ALLSTAR package, we detected all the stars in the 2MASS PSC, with photometry accuracy better that 0.05 mag.
We then performed the  completeness analysis using ADDSTAR routine as mentioned above. The CFs thus obtained for the two 
sub-regions are given in Table \ref{cftable}. The CFs of both the regions seem to be more or less  similar in all the magnitude bins.

 \begin{table}
\caption{Completeness Factors (CFs) of optical and 2MASS $K$-band data within the sub-regions Sh2-252C and NGC 2175s}
\label{cftable}
\begin{tabular}{|p{.75in}|p{.5in}|p{.88in}|p{.89in}|p{1.0in}|} \hline
          & Optical &  &2MASS $K$-band&    \\ \hline
mag. range& Sh2-252C & NGC 2175s &  Sh2-252C  &  NGC 2175s \\
\hline
10 - 11&1.00&1.00&1.00&1.00\\
11 - 12&1.00&1.00&0.99&0.99\\
12 - 13&1.00&1.00&0.97&0.97\\
13 - 14&1.00&1.00&0.92&0.95\\
14 - 15&1.00&1.00&0.85&0.88\\
15 - 16&1.00&1.00&0.67&0.70\\
16 - 17&1.00&1.00&-&-\\
17 - 18&0.99&0.99&-&-\\
18 - 19&0.96&0.98&-&-\\
19 - 20&0.89&0.92&-&-\\
20 - 21&0.80&0.85&-&-\\
21 - 22&0.59&0.71&-&-\\

\hline
\end{tabular}
\end{table}

\section{Results} 

\subsection{Spectral classification of the optically bright sources in Sh2-252}
\label{spectra}

The targets for the  low resolution spectroscopy were selected on the basis of
their brightness, their location within 2 arcmin radius of the sub-regions   
of Sh2-252 (see Fig. \ref{color-opt})  as well as their
location on the optical colour-magnitude diagram (CMD; Fig. \ref{vivcf}).
The coordinates of these sources are given in Table
{\ref{speclog}.   To determine  the  spectral type, we extracted the low-resolution,  
one dimensional   spectrum of each source. The flux calibrated, normalized 
spectrum of   these sources are shown in  Fig. \ref{spec1}. 

In order to classify the stars earlier than B3, we used the criteria given by
Walborn \& Fitzpatrick (1990). For later type stars we used the criteria by Jacoby
et al. (1984) and Torres-Dodgen \& Weaver (1993). Spectra of O and B stars have
the features of hydrogen, helium  and other atomic lines (e.g., \oii,
\ciii, \siiii, \siiv and \mgii etc.). Hydrogen and helium lines are usually seen
in absorption for dwarfs whereas they may be in emission in super giants. In
the case of early type stars, the ratio of {\hei}  4471/{\heii}  4542 is a primary
indicator of the spectral type and the ratio is greater than 1 for spectral type
later than O7. The line strength of {\heii} gets weaker for  late O type stars
and {\heii}  4686 is last seen in B0.5 type stars (Walborn \& Fitzpatrick 1990).
If the spectrum displays {\heii} line at 4200 {\AA}, along with the {\oii}/{\ciii} blend at
4650 {\AA}, it indicates that the spectral type of the star must be earlier than B1. The
absence of {\heii}  4200, {\heii} 4686 and {\mgii}  4481 and the weak features
of silicon lines along with the weak {\oii}/{\ciii} blends at  4070 and 4650 are the
supporting criteria in the B1-B2 range. For type B2, {\hei} is in its maximum
and for later types {\siii}  4128-4130 and {\mgii}  4481 appears stronger (Walborn
\& Fitzpatrick 1990). The presence of the {\hei} lines in absorption constrains the
spectral types to be earlier than B5-B7. Presence of the spectral lines
\mgii  4481 and {\siiii} 4552 is an indication of evolved early B type stars.
The spectral lines such as {\nai}  5893, {\hei}, H$\alpha$, {\cai}  6122, 6162, {\feii}  6456
are used to classify the late type sources (Torres-Dodgen \& Weaver 1993).
Finally, the spectral type was assigned to each source by visual comparison to
the standard library spectra given in the literature (Jacoby et al. 1984) and our
classification is given in Table \ref{speclog}. However, because of low resolution observations
an uncertainty of $\pm$ 0.5 to 1 in the sub-class identification is expected.

Eight out of the fifteen stars observed are found to have spectral types
earlier than B3V. The sub-\hii regions of Sh2-252 (i.e., A, B, C and E) are found
to have at least one star of spectral type early than B3V (i.e., star $\#$1, $\#$3, $\#$5 and
$\#$11). It is quite interesting to note that the remaining four early type sources
(i.e., star $\#$6, $\#$7, $\#$8 and $\#$9)  belong to the small cluster NGC 2175s.
The star $\#$2 shows  H$\alpha$  6563 line in strong  emission (not shown in  Fig. \ref{spec1}). This star also shows
strong NIR excess emission in the $(J - H)/(H - K)$ colour-colour (C-C) diagram  (see  Fig. \ref{jhhk_ob})
most likely originating from the extensive, hot circumstellar material
around the star. The emission line spectra as well as the NIR excess emission
are the primary indicators of an young stellar object (YSO) and hence we tentatively put
this object in the Herbig Be category.

The spectral types of some of these sources have already been mentioned
in the literature.  On the basis of
the 1280 MHz radio continuum flux,  Tej et al. (2006) predicted the
spectral type of the ionizing source of Sh2-252A as B0-B0.5V. The star
$\#$3, located  within 2 arcmin of the radio peak of  Sh2-252A, is the brightest 
source in the optical and NIR wavelengths. This could be the ionizing source
of Sh2-252A. We classify  this source as an  O9.5V type star, which is in 
agreement with the classification by Tej et al. (2006). The
star $\#$5, the probable exciting source of Sh2-252E, had been assigned
a spectral type between  O8 to B0.5V by Garnier \& Lortet-Zuckermann
(1971) and  B1V by Grasdalen \& Carrasco (1975).  Our classification (B1V) 
is in good agrrement  with that given by  Grasdalen \& Carrasco
(1975). The star $\#$11, the brightest source of Sh2-252B  was
classified as B2V by Grasdalen \& Carrasco (1975). However, our
classification (B0V) leads to a class slightly  earlier than their
study. The star $\#$2 had been identified  as a probable
Herbig Be source of spectral type B9  by Chavarr\'{i}a et al. (1989),
which seems to match with our classification.  Thus within  errors the present
spectral classifications are in accordance with those reported in the literature.

\begin{figure}
\centering
\includegraphics[scale = 0.8, trim = 0 0 0 0, clip]{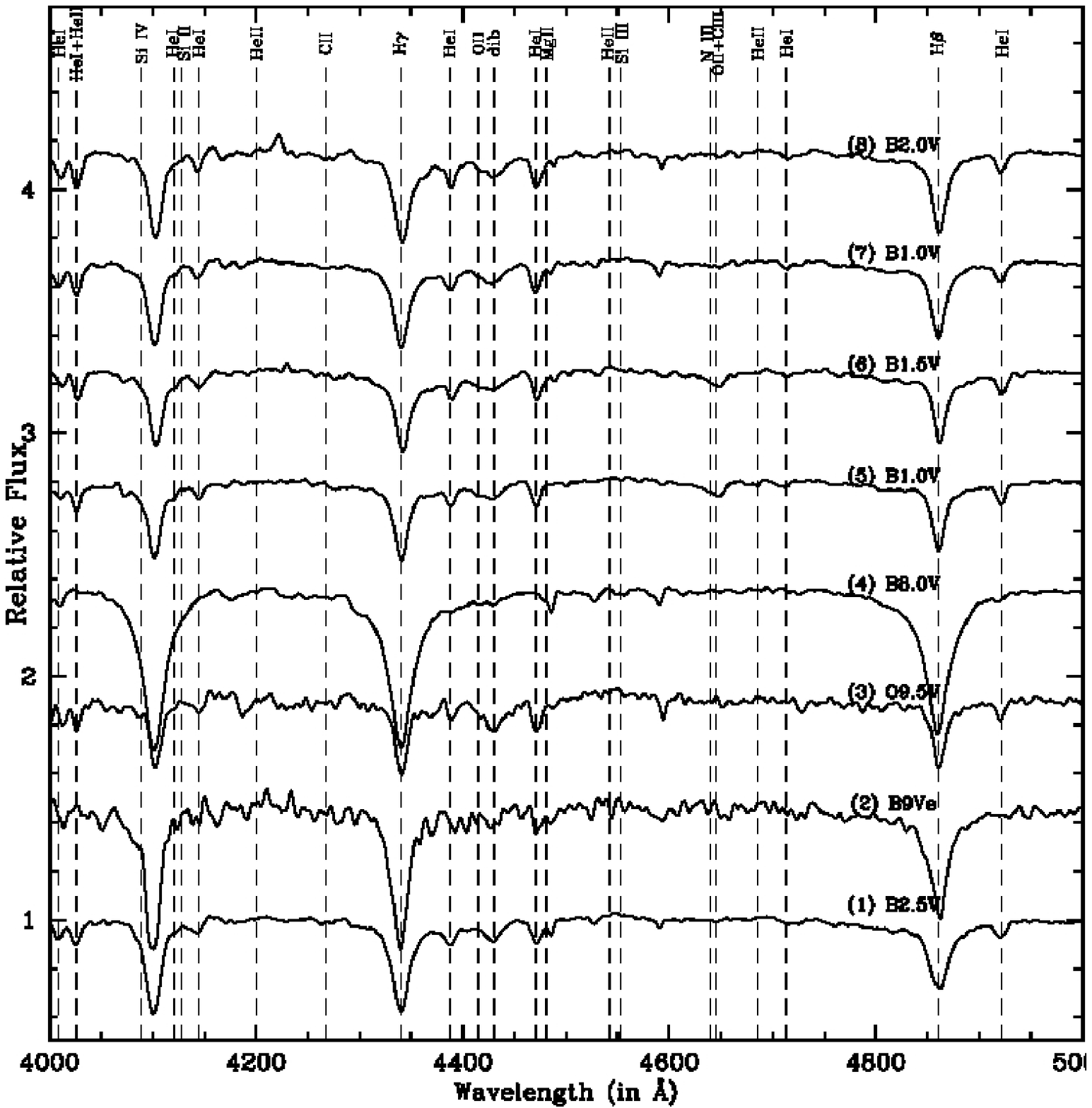}

\caption{Flux calibrated normalized spectra for the optically bright sources in Sh2-252.
The star IDs and spectral classes identified are given in the figure. Important spectral lines are marked. }
\label{spec1}
\end{figure}


\begin{figure}
\centering
\includegraphics[scale = 0.8, trim = 0 0 0 0, clip]{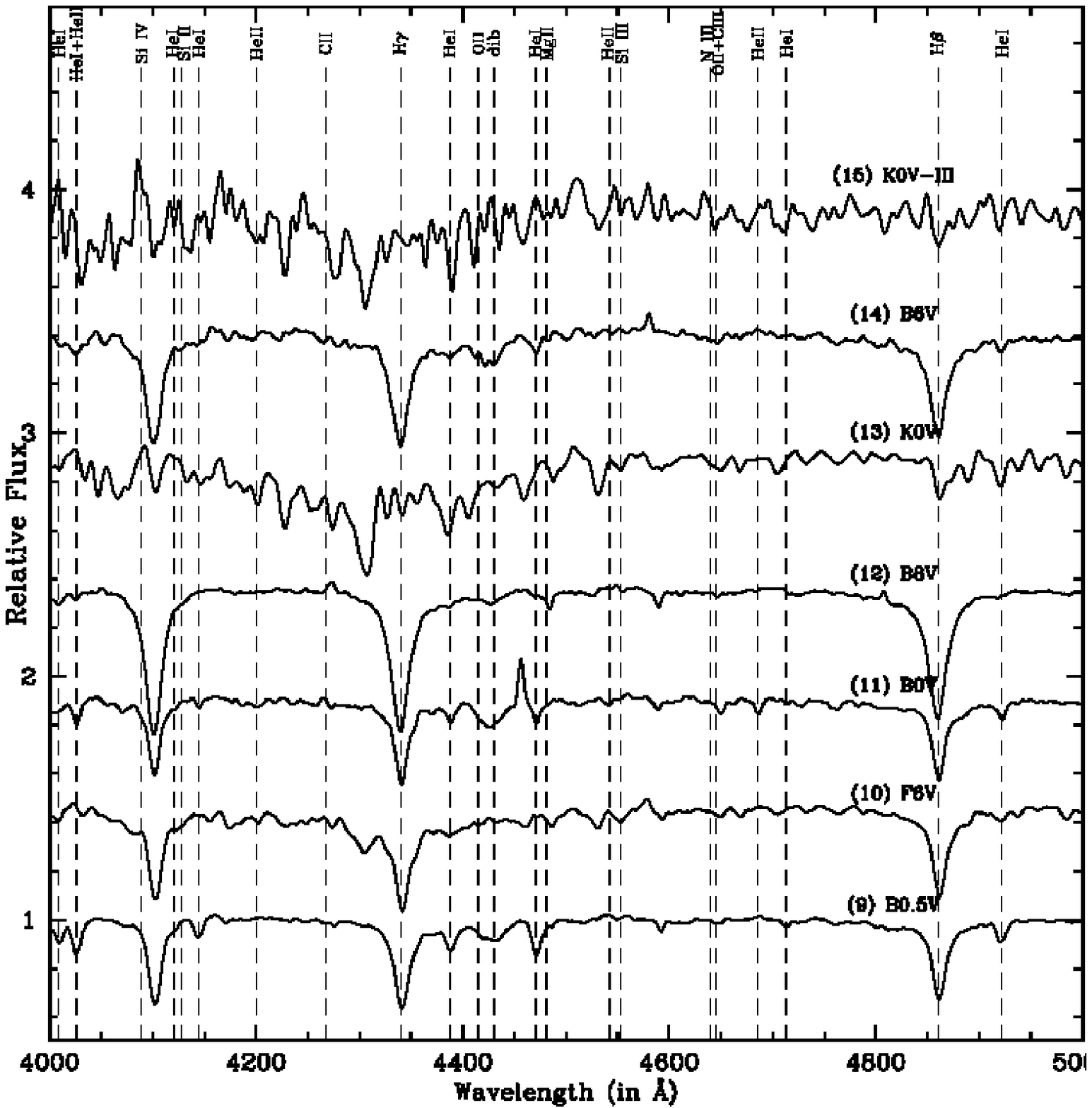}
\noindent{\footnotesize {\footnotesize \hspace{25.0cm} \bf Figure 3 continued.} }

\end{figure}



\begin{table}
\caption{Details of spectroscopically identified stars}
\label{speclog}
\scriptsize
\begin{center}
\begin{tabular}{|p{0.2in}|p{.7in}|p{.8in}|p{.45in}|p{0.4in}|p{0.3in}|p{0.55in}|p{0.5in}|p{0.3in}|p{0.6in}|}
\hline
ID & $\alpha_{(2000)}$ & $\delta_{(2000)}$ & $V$ & $B-V$ & $A_V$ & Spectral & $V_0-M_V$ & D &Associated\\
   &  (h:m:s)        & (d:m:s)          & mag & mag   & mag  &  class    & mag         & pc &with\\
\hline

1 & 06:09:19.51 & +20:38:20.61 &    11.79 &     0.32  & 1.66  & B2.5V & 12.15 & 2690 &C \\
2 & 06:09:21.18 & +20:40:23.01 &    14.11 &     0.90  & -     & B9Ve  & -     & -    &C \\
3 & 06:08:32.10 & +20:39:19.09 &    13.88  &     -      & 6.5   & O9.5V & 11.63 & 2120 &A \\
4 & 06:08:35.31 & +20:39:24.49 &    12.50 &     0.30  & 1.28  & B8V   & 11.47 & 1970 &A \\
5 & 06:09:52.63 & +20:30:16.60 &    11.10 &     0.36  & 1.92  & B1V & 12.37 & 2980 &E \\
6 & 06:10:52.84 & +20:36:42.00 &    11.05 &     -      & 1.93  & B1.5V & 11.94 & 2440 &N2175s \\
7 & 06:10:52.83 & +20:36:34.00 &    10.34 &     -      & 1.67  & B1V   & 11.87 & 2370 &N2175s \\
8 & 06:10:54.10 & +20:36:44.77 &    11.51 &     0.49  & 2.27  & B2V   & 11.69 & 2180 &N2175s\\ 
9 & 06:10:55.97 & +20:36:29.65 &    10.92 &     0.49  & 2.40  & B0.5V & 12.12 & 2660 &N2175s\\
10& 06:10:52.02 & +20:38:00.50 &    10.95 &     0.32  & -     & F6V   & -     & -    &N2175s\\  
11& 06:09:00.14 & +20:38:28.14 &    10.79 &     0.55  & 2.62  & B0V   & 12.17 & 2710 &B\\
12& 06:09:06.49 & +20:39:01.43 &    12.60 &     0.17  & 0.88  & B8V   & 11.97 & 2480 &B\\
13& 06:09:00.65 & +20:35:04.74 &    12.32 &     0.74  & -     & K0V   & -     & -    &B\\
14& 06:11:46.38 & +20:39:21.10 &    13.74 &     -      & 2.77  & B6V   & 11.87 & 2360 &Teu 136\\
15& 06:11:45.10 & +20:39:20.42 &    15.85 &     -      & -     & K0V-III & -     & -    &Teu 136\\
16$^*$& 06:09:39.60 & +20:29:15.43 & 7.55  &     0.06  & 1.14  & O6.5V & 11.74 & 2230 &- \\
\hline
\end{tabular}
\end{center}
For 16$^*$ data is taken from Pismis  (1977)
\end{table}

\subsection{Reddening, membership and distance of the bright sources in Sh2-252}
\label{memb}

\begin{figure}
\centering
\includegraphics[scale = 1, trim = 30 30 100 200, clip]{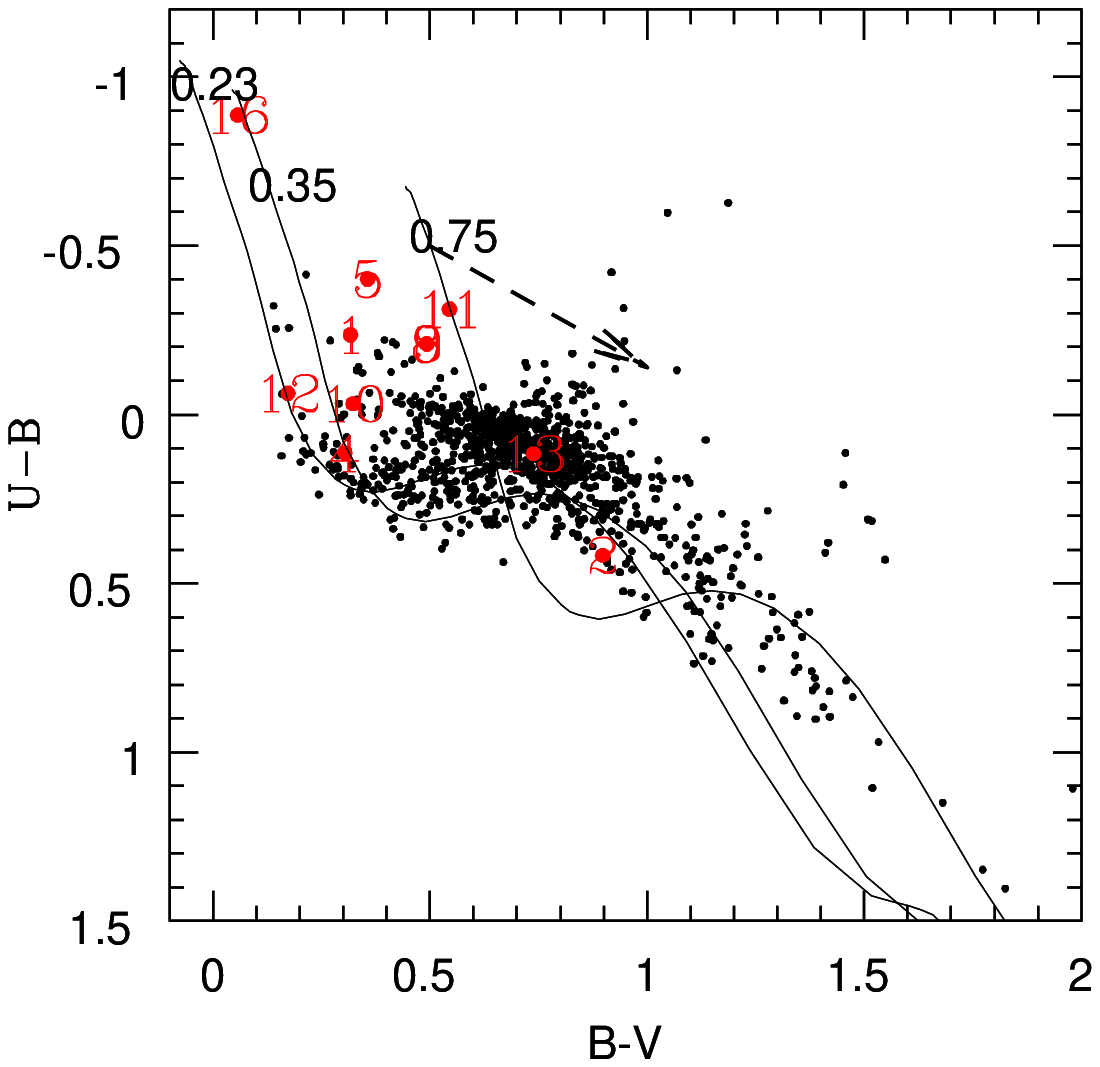}
\caption{$(U  - B)/(B  - V)$ colour-colour  diagram for the stars in Sh2-252 region. The continuous curves 
represent the ZAMS by Girardi et al. (2002) shifted 
along the reddening slope of 0.72 (shown as dashed line) for $E(B-V)$ = 0.23, 0.35 and 0.75 mag,
respectively. The spectroscopically observed sources are shown in red with their IDs 
as given in Fig. \ref{color-opt} and in Table \ref{speclog}.
 }
\label{ubbv}
\end{figure}

The extinction in a star forming region arises due to two distinct
sources: (1) the general ISM in the foreground of the region
[$E(B-−V)_{min}$], and (2) the localized cloud associated with the
region [$E(B−-V)_{local}$ = $E(B-−V) -− E(B-−V)_{min}$], where $E(B−-V )$ is
the observed reddening of the star embedded in the  local cloud. The presence
of the molecular cloud  (Lada \& Wooden 1979) and the nebulosity associated
with  Sh2-252  as evident in Fig. \ref{color-opt} would give rise to
variable extinction  within this region.  To estimate
the extinction towards the region, we used the $(U - B)/(B - V )$
C-C diagram. Fig. \ref{ubbv} shows the $(U - B)/(B - V
)$ C-C distribution of all the stars detected in our  Kiso observations, where
the spectroscopically observed sources are numbered in red. It is to
be noted that some of the spectroscopically observed stars are not covered/detected
in the $U$ and $B$ bands,  hence they are not included in
Fig. \ref{ubbv}. The star $\#$16,  which is the ionizing source of Sh2-252
(i.e., HD 42088), was saturated in our observations. Hence the data
for this star is taken from Pismis (1977).  
The zero-age-main-sequence (ZAMS)  by  Girardi et al. (2002) that is
reddened by $E(B-V)$ = 0.23 mag  along the reddening vector having a
normal slope of $E(U -B)/E(B-V )$  = 0.72  seems to  match with the
first sequence of  distribution in Fig. \ref{ubbv}. However, majority of the 
spectroscopically identified stars earlier than  spectral type B3
show $E(B-V)$ in the range  0.35 - 0.75 mag.  The stars  $\#$10, $\#$13 and 
$\#$15 are found to be late type stars and hence may not be  associated with Sh2-252.
Assuming that uncertainties in the $(U-B)$ and $(B-V )$ colours mainly
originate from their calibration (ref. section \ref{obs}), the typical error in the estimation
of $E(B-V)$ would be $\sim$  0.04 mag.  A careful inspection  of the C-C diagram indicates  the presence of
further reddened population at $(B-V)$ $\sim$ 0.65 - 0.90 mag.  Some of them 
may be the reddened members of the Sh2-252 region. The population having $E(B-V)$ $\sim$ 1.00
- 1.25 mag could be part of the background Norma-Cygnus arm as discussed by
Carraro et al. (2005) and Pandey et al. (2006).

We also estimated the  individual reddening $E(B-V)$ of 
spectroscopically classified stars. We used the relation  $E(B-V)$=
$(B-V) - (B-V)_0$, where $(B-V)$ is the observed colour and $(B-V)_0$
is the intrinsic colour taken from Schmidt-Kaler (1982).  Some of the stars 
do not have the $(B-V)$ data (i.e., stars $\#$3, $\#$6, $\#$7 and $\#$14) and for 
these stars $A_V$ was calculated by using the  relations $E(J-H)$ = 
$(J-H)-(J-H)_0$ and  $A_V$ = $E(J-H)$/0.11 (Cohen et al. 1981), where,
$(J-H)$ colours were taken from  2MASS and intrinsic $(J-H)_0$ colours 
were taken from  Koornneef  (1983). The  $A_V$ thus
calculated are given in Table \ref{speclog}. The range of reddening 
of the spectroscopically identified stars shown in Fig. \ref{ubbv} agree
well  with that given in Table \ref{speclog}.  

\begin{figure}
\centering
\includegraphics[scale = 0.55,  trim = 0 0 0 0,  clip]{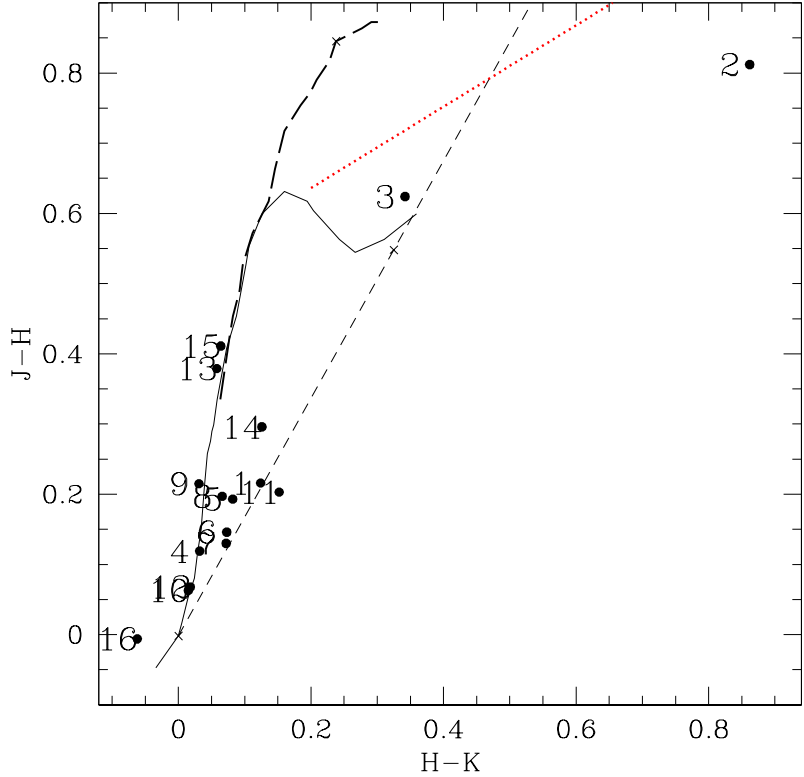}

\caption{$(J-H)/(H-K)$ colour-colour  distribution of the spectroscopically classified stars in Sh2-252. 
Their IDs are as given in Table \ref{speclog}.  The locus for dwarfs (thin solid curve) and giants 
(thick dashed curve) are from Bessell \& Brett (1988). The red dotted  lines represent the 
intrinsic locus of CTTSs (Meyer et al. 1997). The dashed line in black represents the reddening vector drawn from
the early  dwarf locus (Cohen et al. 1981). }
\label{jhhk_ob}
\end{figure}

In Fig. \ref{jhhk_ob} we have shown the $(J-H)/(H-K)$ C-C distribution of all the spectroscopically identified stars in
Sh2-252. The sources falling towards the right of the reddening vector are considered to  have NIR excess (see sect. \ref{nircc}
for details). The star $\#$2 falls in the location where Herbig Ae/Be stars are generally found (Hernandez et al. 2005) and 
this justifies our spectral classification of this source as a Herbig Be type (see sect. \ref{spectra}). 
The figure also shows that star $\#3$ is highly reddened when compared to other massive stars in the field.

The estimated  spectral classification, apparent magnitudes and  extinction 
of the massive members of Sh2-252 (Table \ref{speclog})	 enable us
to derive their distances. Using the  present spectral classification
along with $A_V$ and  $V$ values from Table \ref{speclog} and $M_V$
values from the  table of Schmidt-Kaler (1982), we estimated the
intrinsic distance modulus ($V_0 - M_V$) of each source and the same is given
in Table \ref{speclog}. The average value of the intrinsic distance
modulus  is found to be 11.91 $\pm$ 0.18  mag, which corresponds
to a distance  of $2400^{+220}_{-180}$ pc.   We also calculated the intrinsic distance modulus ($K_0-M_K$)
of each source based on the observed colours from 2MASS data and intrinsic colours and absolute magnitudes
from Koornneef (1983) and Schmidt-Kaler (1982), respectively. The average value of the intrinsic distance 
modulus using 2MASS data was found to be 11.80 $\pm$ 0.10 mag, which is comparable with that of the intrinsic
distance modulus estimated from optical data within errors.   Fig. \ref{bvv0} shows
dereddened $V_0/(B-V)_0$ CMD  for the probable main sequence (MS)  members of Sh2-252 identified
spectroscopically. In  this figure,  we have also plotted  the
theoretical ZAMS  locus  for solar metalicity ($Z=0.02$) by
Girardi et al. (2002), adjusted for the  distance 
of 2.4 kpc, which seems to be matching  well with the
distribution of the probable MS members. The present distance estimate is in good agreement with that of
Georgelin \& Georgelin (1970; 2.4 kpc),  Neckel et al. (1980; 2.3 kpc)
and Chavarr\'{i}a  et al. (1987; 2.3 kpc). At the same time, it is also in agreement with the 
distance  estimates by Grasdalen \& Carrasco (1975; 2.6 kpc)
and Reid et al. (2009;  2.1 $\pm$ 0.03 kpc) within errors.

\begin{figure}
\centering
\includegraphics[scale = 1, trim = 40 80 80  150, clip]{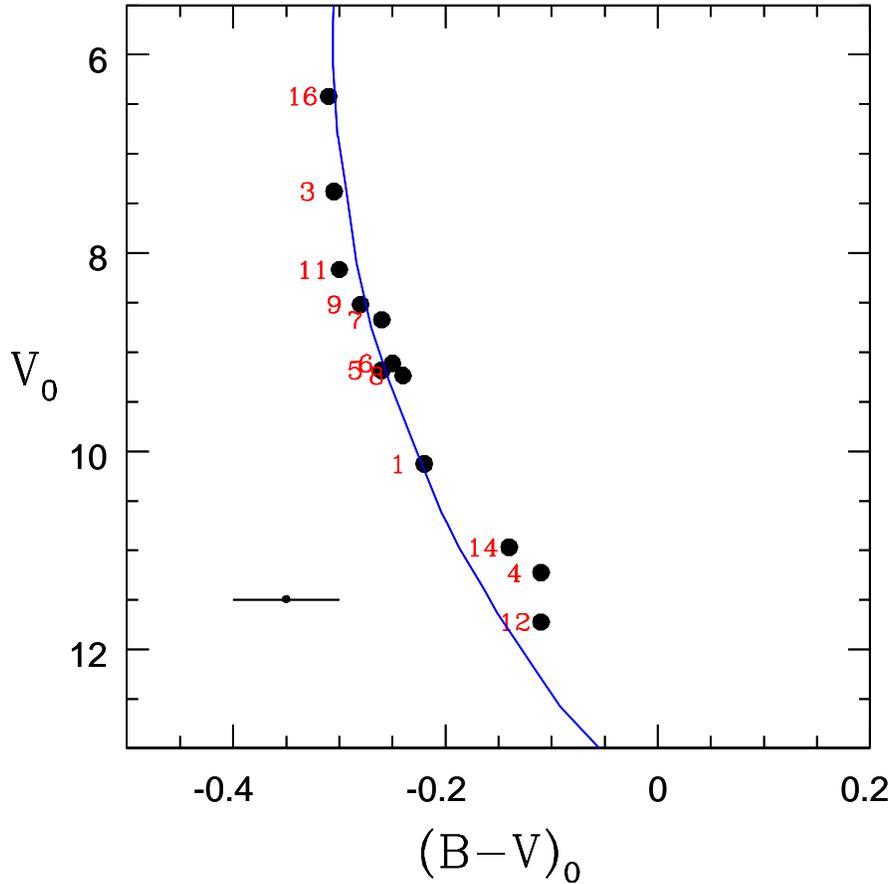}
\caption{$V_0/(B-V)_0$  CMD for the spectroscopically identified massive member stars in Sh2-252. 
The continuous curve represents the ZAMS locus by  Girardi et al. (2002) corrected for the  distance of 2.4 kpc.
The sources are  numbered as in Table \ref{speclog}. The average error in  colour is shown in the lower left side 
of the figure.
}
\label{bvv0}
\end{figure}

An important issue to be discussed here is  the association of
NGC 2175s and Teu  136 with Sh2-252. NGC 2175s is a small, optically visible
cluster located at the east of Sh2-252 (at $\alpha_{2000}$ =
$06^{h}10^{m}52^{s}$;  $\delta_{2000}$ =
$+20^{\circ}36^{\prime}36^{\prime\prime}$; see Figs. \ref{area} and
\ref{color-opt}). The photoelectric study by Pismis (1970) put this
cluster at a distance of 3500 pc. However, in the present  spectroscopic
observations, we  identified four early B type   members of this cluster (i.e.,
stars $\#$6, $\#$7, $\#$8 and $\#$9, see Table \ref{speclog}) and the distance
estimates of these four  sources are in good agreement with that of
the average distance of Sh2-252.  Hence, our spectroscopic analysis
supports the  association of  NGC 2175s with the Sh2-252 complex. 
Similarly, the  cluster  Teu 136 was identified as a NIR embedded cluster by Koposov et al. (2008)
in a survey of the 2MASS catalog designed to identify new open clusters.
Located at coordinates $\alpha_{2000}$ = $06^{h}11^{m}56^{s}$;  $\delta_{2000}$ =
$+20^{\circ}40^{\prime}14^{\prime\prime}$, $\sim$ 15$^{\prime}$ NE of NGC 2175s, this cluster lies close to
the eastern edge of the Sh2-252 complex.  We obtained spectra of two  stars located close to this 
cluster (i.e., stars $\#$14 and $\#$15) and the distance of star $\#$14 (see Table \ref{speclog}) seems to 
match well with that of Sh2-252.  Also, the spectral types of  two bright stars at the center of this cluster, which are 
saturated in our photometric observations are given as B1.5 and B2.5, respectively in Reed (2003). The average distance
modulus of these two stars comes out to be 11.74 mag which seems to  match with the average distance 
modulus of the Sh2-252 complex.  
Therefore we presume that Teu 136 could be associated with the Sh2-252 complex, too.  
Hence, in this paper we report that the NIR cluster Teu 136 as a sub-cluster of the Sh2-252 complex. 
 
\begin{figure}
\centering
\includegraphics[scale = 0.85, trim = 0 10 0 20, clip]{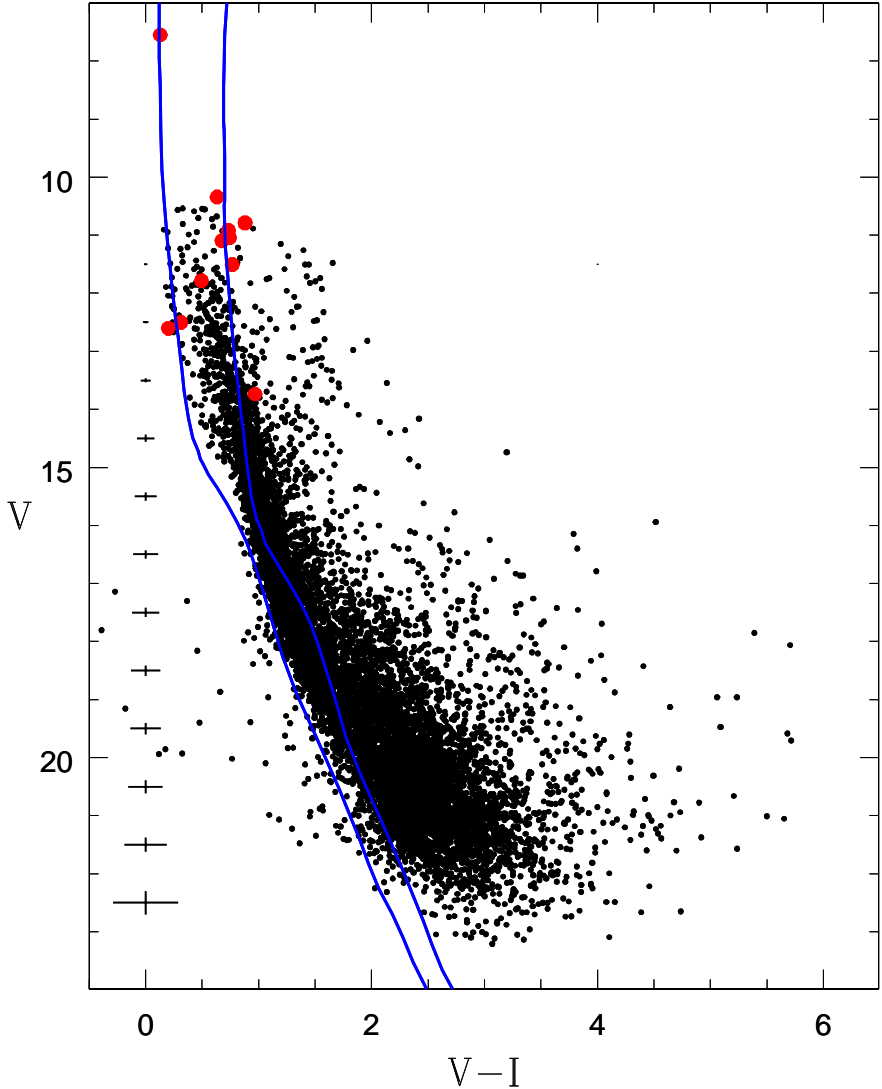}
\caption{$V/(V-I)$ CMD for all the stars detected in our optical photometry. The dashed curves 
are the  ZAMS from Girardi et al. (2002)  corrected for the  distance 2.4 kpc and reddening  $E(B-V)$ = 0.35 
and $E(B-V)$=0.75, respectively.  The sources marked with red filled circles 
are the spectroscopically identified  MS members of Sh2-252.  The average error in $(V-I)$  colour as a function of magnitude 
is shown in the  left side  of the figure.
}
\label{vivcf}
\end{figure}

In Fig. \ref{vivcf}, we show the $V/(V - I)$ CMD for all the stars
detected in our optical photometry. Spectroscopically identified  massive MS
members  are shown with large filled circles in red,  where the brightest star is
HD 42088. The dashed curves
are the ZAMS by Girardi et al. (2002) shifted for  the 
distance of 2.4 kpc and  reddening $E(B-V)$=0.35 and $E(B-V)$=0.75, respectively. 
 The average photometric uncertainty in $V$ magnitudes and $(V-I)$ colours estimated from the difference between the 
input and recovered magnitudes of the artificial stars (see Sect. \ref{cf}) are shown in the left side of the figure.
The CMD clearly shows a broad  MS  band down to $V$ $\sim$ 14 mag with a significant number of sources located 
within the reddening strips (i.e.,  $E(B-V)$ = 0.35 to 0.75 mag). In the absence of spectroscopic information, 
it is difficult to assign membership for these bright sources. Since
Sh2-252 is  located close to  the Galactic plane (l=190.04;  b=+0.48)
at a distance of 2.4 kpc, it is quite  natural that the  region could
be significantly contaminated by Galactic field MS and giant populations.
The extinction variation inside the Sh2-252 region
may cause a spread in the MS both for the member sources as well as
for the background sources. A significant number of sources can also be noticed
towards the right side of the ZAMS, part of which could be PMS stars.

\subsection{Age and mass distribution of candidate YSOs in Sh2-252}

In the absence of proper motion data, the  membership of  Sh2-252  can be established  by identifying
stars with indicators of youth. YSOs emit excess radiation in the infrared in comparison to 
MS stars due to the thermal emission from their circumstellar material. So  YSOs can be 
identified on the basis of their  infrared excess emission. We have also carried out a YSO survey of the 
Sh2-252 region using the 2MASS  and the {\it Spitzer}-IRAC data sets (3.6, 4.5, 5.8 and 8.0 $\mu$m) 
and the detailed analyses will be presented 
in a forthcoming paper. A total of 96 Class I and 400 Class II YSOs are identified within this
complex on the basis of IR colour excess criteria given by Gutermuth et al.
(2009). The possible contamination such as PAH-emitting galaxies, broad-line
AGNs etc. have been excluded using the criteria given by Gutermuth et al.
(2009). Similarly, the presence of H$\alpha$ emission line is considered as a significant
characteristic of a YSO with ongoing disk accretion process (e.g., Dahm  2005). We identified 61 H$\alpha$  emission
line stars in our slitless spectroscopy survey of the  Sh2-252 region (see section \ref{slitless}).
Our deep photometry provided $V$ and $I$ counterparts for 15 Class I, 179 Class II
and 54 H$\alpha$  emission line sources.  Of these 54 H$\alpha$  sources 4 are Class I and 33 are Class II YSOs. 
The distribution of these candidate YSOs i.e., Class
I, Class II and H$\alpha$  emission line sources on the $V/(V-I)$ CMD is shown in Fig. \ref{viv} by using red
squares, green triangles and blue triangles, respectively. The distance to Sh2-252 is adopted
as 2.4 kpc as estimated on the basis of the spectroscopic analyses mentioned in section \ref{memb}. In
Fig. \ref{viv} the ZAMS (thick solid curve) by Girardi et al. (2002) and the PMS isochrones (dashed curves)
by Siess et al. (2000) for age 0.1 and 5 Myr are also shown. The ZAMS and isochrones are
shifted for the distance of 2.4 kpc and average reddening i.e., $E(B-V)$ = 0.5 mag.
The average photometric uncertainty in $V$ magnitudes and $(V-I)$ colours estimated from the completeness simulation (see Sect. \ref{cf}) 
close to region C are shown in the right  side of the figure.   The
distribution of YSOs on the $V/(V-I)$ CMD is an ideal tool to estimate the approximate ages
of YSOs. It is evident from this figure that a majority of the YSOs in Sh2-252 are distributed
between the PMS isochrones of age 0.1 and 5 Myr, which shows that the region Sh2-252 is
significantly comprised of young sources.  However, the photometric uncertainty, differential
reddening, binarity, different evolutionary stages of YSOs etc. can cause a spread in the CMD. 
The spectroscopic analysis of the  member stars shows that the reddening $A_V$ in the region varies between  
0.9 to 6.5 mag (see Table \ref{speclog}). Also,  the NIR colour-colour analysis of the candidate PMS sources  shows 
that they are reddened up to 8 mag (see Sect. \ref{nircc}), which suggest that the
region is significantly affected by the differential reddening.  Similarly, the photometric uncertainties in the case of YSOs (see Fig. \ref{viv}) 
may be relatively large  due to crowding and variable background which  may cause the scatter in the distribution of the Class I and Class II 
sources in the CMD. Hence the  differential reddening and photometric uncertainty would have an impact on the observed age spread in the CMD. 
The detailed spatial distribution and evolutionary status of the YSOs in the Sh2-252 region will be analyzed
in the forthcoming paper. In Fig. \ref{viv} we have also shown the PMS evolutionary tracks by
Siess et al. (2000) (thin solid curves) for various mass bins (the value of mass for each track
is given towards its right), which indicate that a majority of the YSOs have masses in
the range between 0.3 $M_{\odot}$ - 2.5 $M_{\odot}$.

\begin{figure}
\centering
\includegraphics[scale = .55, trim = 0 10 0 10, clip]{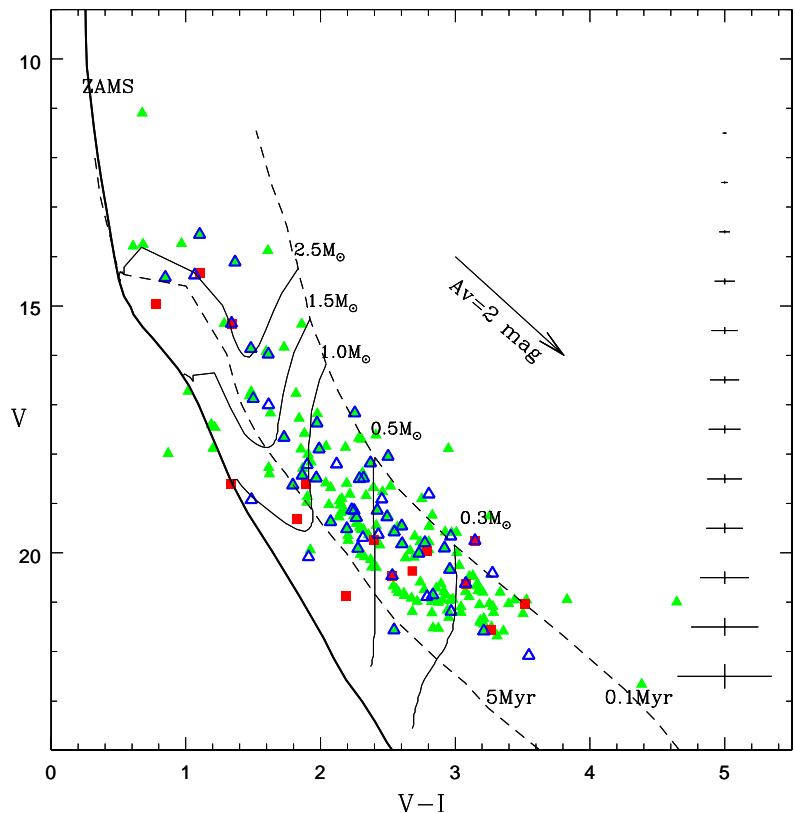}
\caption{$V/(V-I)$ CMD for the identified candidate YSOs in Sh2-252 within the area shown by solid lines in 
Fig. \ref{area}. The Class I and Class II sources are shown with red squares and green filled triangles, 
respectively and
the blue open triangles are the $H\alpha$ emission line sources. The thick solid curve is the 
locus of ZAMS from Girardi et al. (2002), dashed curves are the PMS isochrones of age 0.1 and 5 Myr, 
respectively and the thin solid curves are the evolutionary tracks for various mass bins from 
Siess et al. (2000).  All  the isochrones and tracks  are corrected for  the  distance  and
reddening. The arrow indicates  the reddening vector for $A_V$ = 2 mag.  The average error in $(V-I)$  colour as a function of magnitude 
is shown in the  right side  of the figure.
 }
\label{viv}
\end{figure}

The ages of the young clusters/\hii regions can also be derived from the
post-main-sequence age of their most massive members. The most massive  member of
Sh2-252 is HD 42088, an O6.5 star which is still in the MS (Conti \& Alschular
1971; Walborn 1972) and is considered to be the main ionizing source of
Sh2-252. Also, none of the spectroscopically identified OB stars of Sh2-252 are found to
be evolved (see section \ref{spectra}). Hence, the age of the \hii region Sh2-252 should be younger or
of the order of the MS life time of the O6.5V star. i.e., $\sim$ 4.4 Myr (Meynet et
al. 1994). Thus we can put an upper limit to the age of Sh2-252 as $\sim$ 4 Myr.
However, this has to be taken as an approximate estimate as the low mass
members of the region could have been formed prior to HD 42088.

\subsection{Stellar contents of the sub-regions in Sh2-252}

\subsubsection{Ionizing sources of the C\hii regions}
\label{isource}

As discussed in section \ref{overview}, the radio observation of this region by Felli et al.
(1977) identified six compact thermal radio sources, of which the regions A, B,
C and E have been claimed as C\hii regions having its own ionizing sources in
each of them. In order to search for the ionizing sources of these regions, we
obtained spectra of optically bright sources within 2$^\prime$ radius around each region.
The spectral types of these bright sources have been discussed in section \ref{spectra}
and each region is found to have at least one source with spectral type earlier
than  B3V. Their spectral type can also be obtained by measuring the number of
Lyman continuum photons (N$_{lyc}$ ) emitted per second by the star that would
be responsible for the ionization of each region. From the radio continuum flux
at 1415 MHz by Felli et al. (1977) together with our estimated distance of 2.4
kpc and assuming an electron temperature of 10$^4$ K, we calculated log N$_{lyc}$ using
the relation given by Mart\'{i}n-Hern\'{a}ndez et al. (2003). The log N$_{lyc}$ have been
obtained as 46.73, 47.52, 46.57 and 46.55, respectively for regions A, B, C and
E. From these log N$_{lyc}$ values  we derived the spectral type of the ionizing sources (Panagia
1973) as  B0-B0.5, B0, B0.5 and B0.5, respectively,
for regions A, B, C and E. These spectral types  are consistent with
the earliest spectral types derived by our optical observations for regions A, B
and E (i.e., star $\#$3, $\#$11 and $\#$5, see Table \ref{speclog}) within the errors. The 1280 MHz
observations by Tej et al. (2006) also estimated the spectral type of the ionizing
source of region A as  B0-B0.5V which is consistent with our estimate. Slight
difference in the spectral type estimate is expected due to uncertainties in the
effective temperature and ionizing flux of massive stars, for a given spectral
type. Thus we conclude that the stars $\#$3, $\#$11 and $\#$5 are the ionizing sources
of the C\hii regions A, B and E, respectively. However in the present observation, the earliest spectral type
derived in region C is  B2.5V (see Table \ref{speclog}), which does not seem to be
early enough to explain the radio continuum flux. The expected spectral
type from the radio flux is much earlier than B2.5V. The study by K\"{o}mpe
et al. (1989) showed that the probable spectral type of the  IRAS point source 
(IRAS 06063+2040) located at the center of the nebulosity in region C is B0V, which could
be the candidate ionizing source of this C\hii region. Since this source is deeply
embedded in the cloud  ($K$=9.16, $A_V$ $\sim$ 8 mag), we could not carry out optical spectroscopy of this source 
and future spectroscopy  would help us to confirm this result.  In Fig. \ref{cont} we have shown the $K_S$-band images of 
the sub regions A, B, C, E, NGC 2175s and Teu 136 for an area 8$^\prime$ $\times$ 8$^\prime$. The probable
ionizing sources of the regions A, B, C and E are marked with a white star symbol.

\subsubsection{Stellar surface density distribution in Sh2-252}
\label{sdensity}

\begin{figure}
\centering
\includegraphics[scale = 0.42, trim = 40 60 40 60, clip]{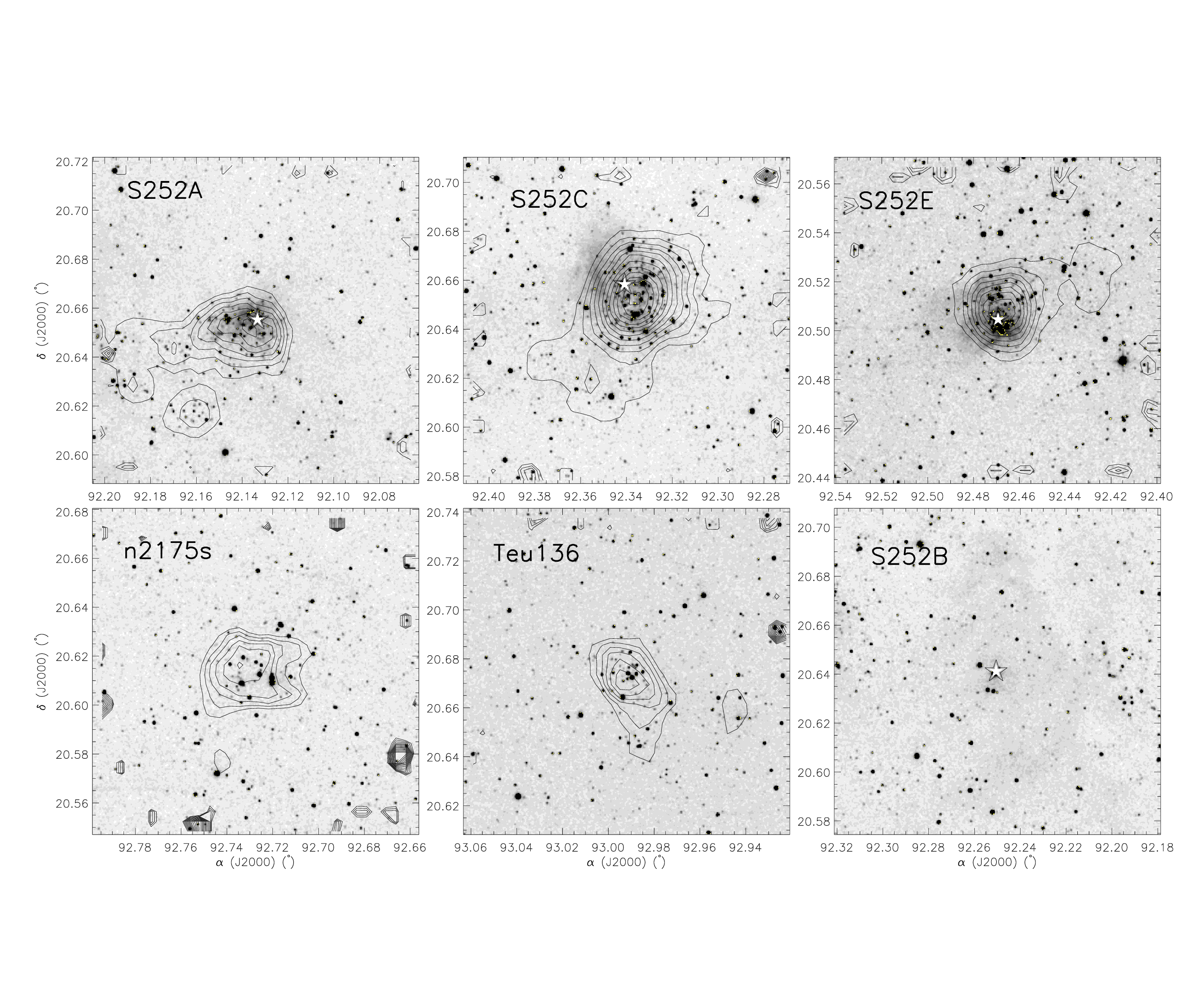}
\caption{8$^\prime$ $\times$8$^\prime$ $K_S$-band images of the sub-regions within Sh2-252. 
The contours are the  stellar surface  density distribution generated 
by using all the sources detected  in $K_S$-band. The ionizing sources of the C\hii regions are also 
marked using white star symbol. }
\label{cont}
\end{figure}

Though we have the optical and NIR observations for an area  $\sim$  1 degree $\times$ 1 degree around the
Sh2-252 region, however a majority of the members of Sh2-252 might be confined within its
sub-regions (see section \ref{overview}). In order to define the extent of each sub-region and to study the
morphology of the region, we generated isodensity contours for stellar population detected
in 2MASS $K$-band.   The two dimensional stellar surface density distribution of the Sh2-252 region
manifests prominent stellar density enhancements at the locations of A, C, E, NGC 2175s and Teu 136.
This indicates the presence of clustering associated with these sub-regions. On the other hand, no clustering is 
apparent towards the sub-regions B and F. In Fig. \ref{cont} 
the stellar surface density contours above  3$\sigma$  of the background level 
are over plotted on the $K_S$-band  images of regions A, B (no contours although), C, E, NGC 2175s and Teu 136 . 
We will discuss the properties of these  five sub-regions i.e., A, C, E, NGC 2175s
and Teu 136 individually in ensuing sections. The extent of these sub-regions
is estimated on the basis of the surface density distribution. The contour above the 
3$\sigma$ background level around each region is considered as its extent. Thus we estimated the radius 
as 3$^\prime$.5, 3$^\prime$.5, 3$^\prime$.0,
3$^\prime$.0, 3$^\prime$.0, respectively, for the region A, C, E, NGC 2175s and Teu 136. We used
these extents for further analysis of these sub-regions. 

\subsubsection{NIR colour-colour diagrams}
\label{nircc}

NIR C-C diagrams are  ideal tools to identify the candidate PMS sources  having NIR 
excess (Hunter et al. 1995; Haisch et al. 2000; 2001; Sugitani  et al. 2002; Devine
et al. 2008; Chavarria et al. 2010).  The $(J-H)$/$(H-K)$ C-C diagrams for  the sub-regions 
 A, C, E, NGC 2175s and Teu 136  and a nearby control field  are  shown in   
Fig. \ref{jhhk}. We have  considered an area of  3$^{\prime}$.0 radius  centered at  $\alpha_{2000}$ =
$06^{h}09^{m}30^{s}$, $\delta_{2000}$ = $+20^{\circ}08^{\prime}$ as
the control field.  This region is  devoid of nebulosity in the $R$-band 
(see Fig. \ref{area}), hence can be  considered as a  field region.  
The  thin and thick solid curves are the locations of unreddened MS and giant stars
(Bessell $\&$ Brett 1988), respectively. The dotted  lines represent the locus 
of intrinsic and reddened ($A_V$ = 4.0 mag; 8.0 mag) classical T Tauri stars (CTTSs; Meyer et al. 1997). 
The two long parallel  dashed lines are the reddening vectors for the early MS and giant type  stars 
(drawn from the base and tip of the two branches). One more reddening vector is plotted from the 
tip of the unreddened CTTS locus. The crosses on the reddening vectors are separated by  an
$A_{V}$ value of  5 mag. The extinction ratios, $A_J/A_V = 0.265, A_H/A_V = 0.155$ and $A_K/A_V=0.090$, 
are adopted from  Cohen et al. (1981).  The magnitudes, colours of the stars and the curves are in the CIT system. 
\begin{figure}
\centering
\includegraphics[scale = 0.9,  trim = 0 0 0 0,  clip]{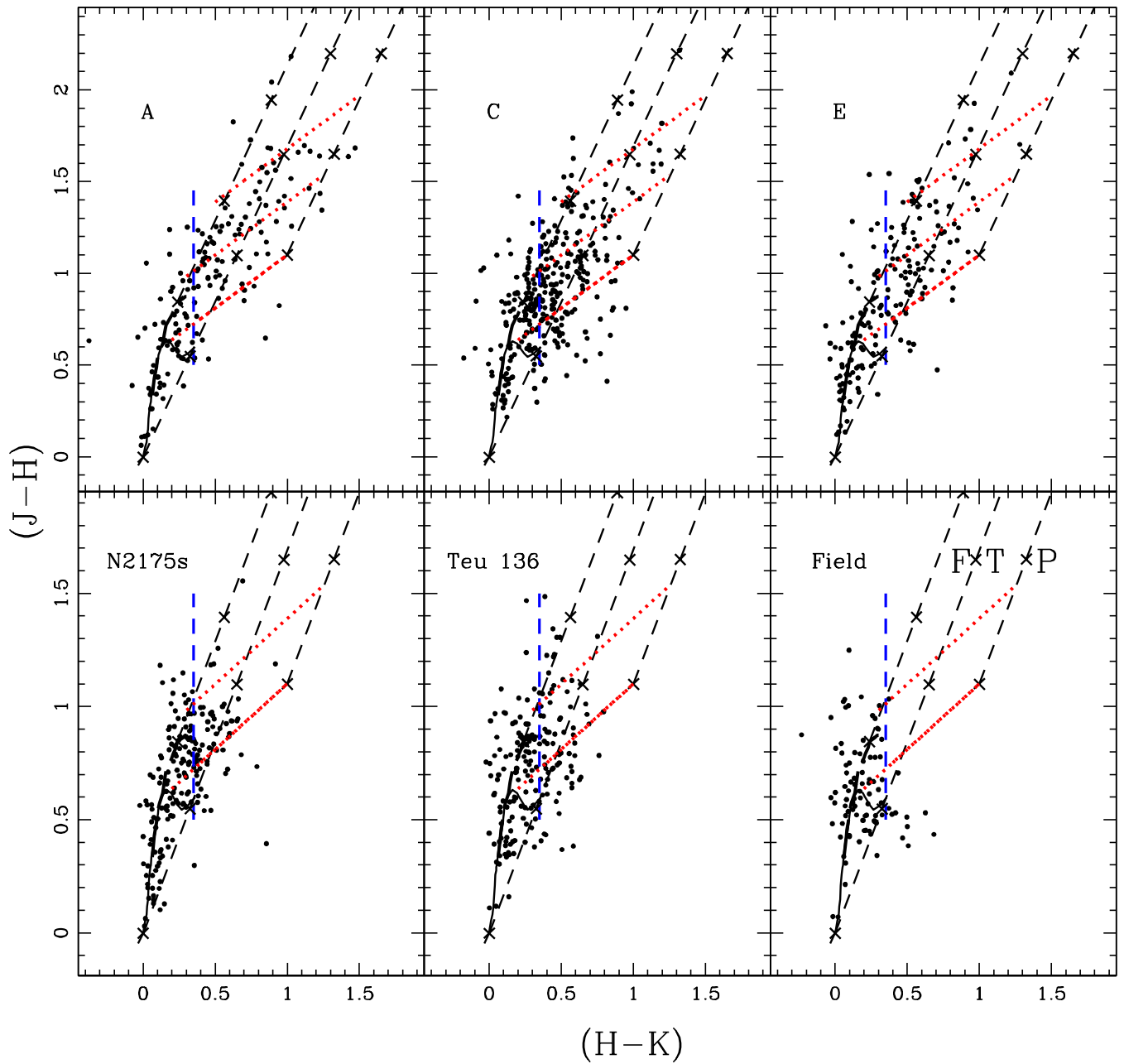}

\caption{$(J-H)/(H-K)$ C-C diagrams of the sources detected in the $JHK$-bands within the sub-regions of Sh2-252 and a nearby
control field. The regions are marked in the figure. The locus for dwarfs (thin solid curve) and giants 
(thick solid curve) are from Bessell \& Brett (1988). The red dotted  lines represent the 
intrinsic and reddened ($A_V$ = 4.0 mag; 8.0 mag) locus of CTTSs (Meyer et al. 1997). The dashed lines 
in black represent the reddening vectors (Cohen et al. 1981). The crosses on the dashed lines are separated by $A_V$ = 5 mag. 
The blue vertical line divides the $(H-K)$ colour  at 0.35 mag (see the text).}
\label{jhhk}
\end{figure}

Following Ojha et al. (2004a), we classified sources according to their locations in $(J-H)/(H-K)$ C-C diagrams.
The `F' sources are those located between  the reddening vectors projected from the intrinsic
colours of MS and giant stars. These sources are  reddened field stars (MS and  giants) or Class
III/Class II sources with little or no  NIR excess (viz., weak-lined T Tauri sources (WTTSs)
but some CTTSs may also be included). The sources located redward of region `F' are considered
to have NIR excess. Among these, the `T' sources are located redward of `F' but blueward of the
reddening line projected from the red end of the CTTS locus. These sources are considered to be
mostly CTTSs (Class II objects)  with large NIR excesses (Lada \& Adams 1992). There may be an
overlap in NIR colours of Herbig Ae/Be stars and T Tauri stars in the `T' region (Hillenbrand et
al. 1992). The `P' sources are those located in the region redward of region `T' and are most
likely Class I objects (protostellar-like) showing large amount of NIR excess. Here it is worthwhile
to mention that Robitaille et al. (2006) have shown that there is a significant overlap between
protostellar-like objects and CTTSs in the C-C diagram.

A comparison of the distribution of sources in the sub-regions and the control field in Fig. \ref{jhhk}
suggests that there is an appreciable difference between them. A significant fraction of sources
within the sub-regions are concentrated between the intrinsic and reddened CTTS locus
(i.e., within the `F' and `T' regions), whereas a majority of the sources in the control field are
mainly concentrated towards the left of the `F' region (i.e., $(H-K) <$ 0.35 mag).  Average value of
the uncertainty in $(H-K)$ colour is $\sim$ 0.05 mag. Taking this uncertainty into consideration and on the basis
of the comparison of the C-C diagrams, we can safely assume that a  majority of the sources in the
sub-regions located between the intrinsic and reddened CTTS locus with $(H-K) >$
0.40 mag and lying in the `F', `T' and `P' regions are most likely PMS members. Some of the above sources lying in the `F' region
could be the reddened field stars, however a majority of them are likely candidate WTTSs
or CTTSs with little or no NIR excess.  Fig. \ref{jhhk} shows that the clusters NGC 2175s
and Teu 136 do not have many PMS members with $A_V >$ 4.0 mag, whereas a significant
fraction of the PMS members of the sub-regions A, C and E  are reddened up to 8.0 mag. 
This indicates that the clusters NGC 2175s and Teu 136, which are located towards the east of
Sh2-252, are less reddened in comparison to other sub-regions of Sh2-252 located
towards the west.

\subsubsection{Optical colour-magnitude diagrams}
\label{optcc}

The $V/(V-I)$ CMDs for the sources within the sub-regions of Sh2-252 and the nearby control
field are shown in Fig. \ref{vivall}. The ZAMS (thick solid curve) by Girardi et al. (2002) and PMS
isochrones (dashed curves) by Siess et al. (2000) for age 0.1 and 5 Myr are also shown.
The ZAMS and isochrones are shifted for the distance of 2.4 kpc and  reddening. 
The encircled sources are those located above the intrinsic 
CTTS locus in the $(J-H)/(H-K)$ C-C diagram (Fig. \ref{jhhk}) and with  $(H-K) >$ 0.4 mag. 
A majority of these sources are found to have ages $<$ 5 Myr in all the sub-regions, suggesting
that the sources with $(H-K)>$ 0.4 mag could be probable PMS members
as discussed in section \ref{nircc}. A comparison of these CMDs with that of the 
control field also supports the notion that the sources lying between 0.1 and 5 Myr
isochrones could be the PMS sources associated with the sub-regions. Similarly,
the CMDs of the sub-regions  resemble  the $V/(V-I)$ CMD for
the candidate YSOs shown in Fig. \ref{viv}. It further supports that the majority of the sources
identified on the basis $(J-H)/(H-K)$ C-C diagrams (Fig. \ref{jhhk}) in the sub-regions
could be the PMS members and that they have an age spread of  5 Myr.  A detailed analysis on the
evolutionary status of the YSOs within the individual regions will be done in the forthcoming paper.

\begin{figure*}
\centering
\includegraphics[scale = .8, trim = 0 0 0 0, clip]{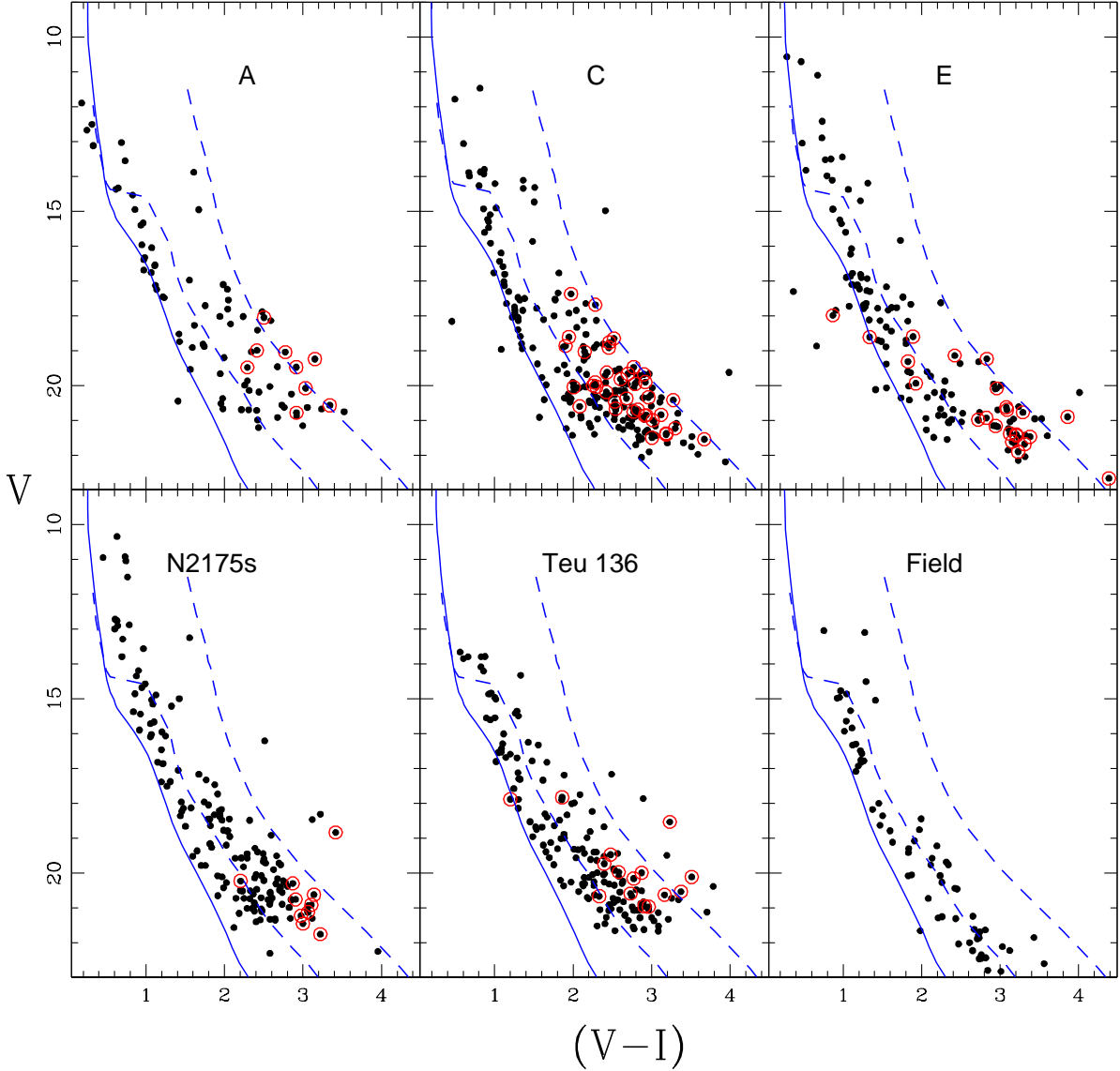}
\caption{$V/(V-I)$ CMDs for the sources within the sub-regions of Sh2-252 and the nearby control field. 
The sub-regions are marked in the figure. Encircled are the sources with $(H-K) >$ 0.4 mag and are located above the intrinsic 
CTTS locus of the $(J-H)/(H-K)$ C-C diagram (Fig. \ref{jhhk}). 
The thick solid curve is the ZAMS from Girardi et al. (2002) and the dashed curves are the PMS 
isochrones of age 0.1 and 5 Myr, respectively, from Siess et al. (2000), corrected
for the distance and  reddening.}
\label{vivall}
\end{figure*}

\subsection{$K$-band luminosity functions of sub-regions in Sh2-252}
\label{klfs}

The luminosity function in the $K$- band is frequently used in studies of young clusters and
star forming regions as a diagnostic tool of the mass function and the star formation history
of their stellar populations. Pioneering work on the interpretation of KLF was presented
by Zinnecker et al. (1993). During the last decade several studies have been carried out
with the aim of determining the KLF of young clusters (e.g., Muench et al. 2000; Lada \&
Lada 2003; Ojha et al. 2004b; Sanchawala et al. 2007; Pandey et al. 2008; Jose et al. 2008;
2011). We used the 2MASS $K$- band data to study the KLF of the Sh2-252 region. Because this
region shows clusterings around the regions A, C, E, NGC 2175s and Teu 136 (see
section \ref{sdensity}), we estimated the KLF for each region independently. The KLF is calculated
within the area of each region as estimated on the basis of stellar surface density distribution
discussed in section \ref{sdensity}.

KLF is the number of stars as a function of $K$- band magnitude. In order to convert the
observed KLF to the true KLF, it is necessary to account for the data incompleteness as well
as the background and foreground source contamination. 
 The regions C and NGC 2175s were corrected for their respective CFs given in Table \ref{cftable}
and the average value of the CFs for regions C and NGC 2175s  was applied to  regions A, E and Teu 136 
to  correct for their data incompleteness.
The data from the control field region (see section \ref{nircc}) has been used to remove the
field star contribution. Since this control field is located off the Sh2-252 region, its
background population will be affected by a smaller  interstellar reddening. Whereas,
the background field stars of the  sub-regions of Sh2-252 are seen
through a larger reddening due to the matter associated with it.
Hence the population in the background of sub-regions will be more reddened as
compared to the control field. Therefore  a direct subtraction of the observed control
field from the sub-regions of Sh2-252 will yield incorrect LF. To account for the
higher extinction towards the embedded sub-regions, the field star population towards the
direction of the control field is predicted on the basis of  the Besan\c con  Galactic model of stellar
population synthesis (Robin et al. 2003) by using a similar procedure as described by Ojha et
al. (2004b). An advantage of using this model is that we can simulate foreground (d $<$ 2.4
kpc) and the background (d $>$ 2.4 kpc) field star populations separately. The foreground
population was simulated by using the model with the  extinction $A_V$ = 1.1 mag
($E(B-V)$ = 0.35 mag; cf. section \ref{memb}) and d $<$ 2.4 kpc.  As discussed in section \ref{nircc},
a majority of the PMS members of regions  A, C and E are reddened up to $A_V$=8.0 mag, whereas in NGC 2175s and Teu 136
they are reddened up to $A_V$=4.0 mag. Hence the background population in these regions are seen through  clouds with
extinction of  8.0 and 4.0 magnitudes, respectively. 
Hence, we simulated two sets of background population
(d $>$ 2.4 kpc) with $A_V$ values of 4.0 and 8.0 mag, respectively. Then we determined the
fractions of the contaminating stars (foreground $+$ background) over the total model counts.
The fraction was calculated separately for both sets of the simulated background population.
This fractions were used to scale the nearby observed control field and subsequently the star
counts of the modified control field were subtracted from the KLFs of the sub-regions of
Sh2-252 to obtain their final corrected KLFs.

\begin{figure}
\centering 
\includegraphics[scale = 0.8, trim = 0 0 0 0, clip]{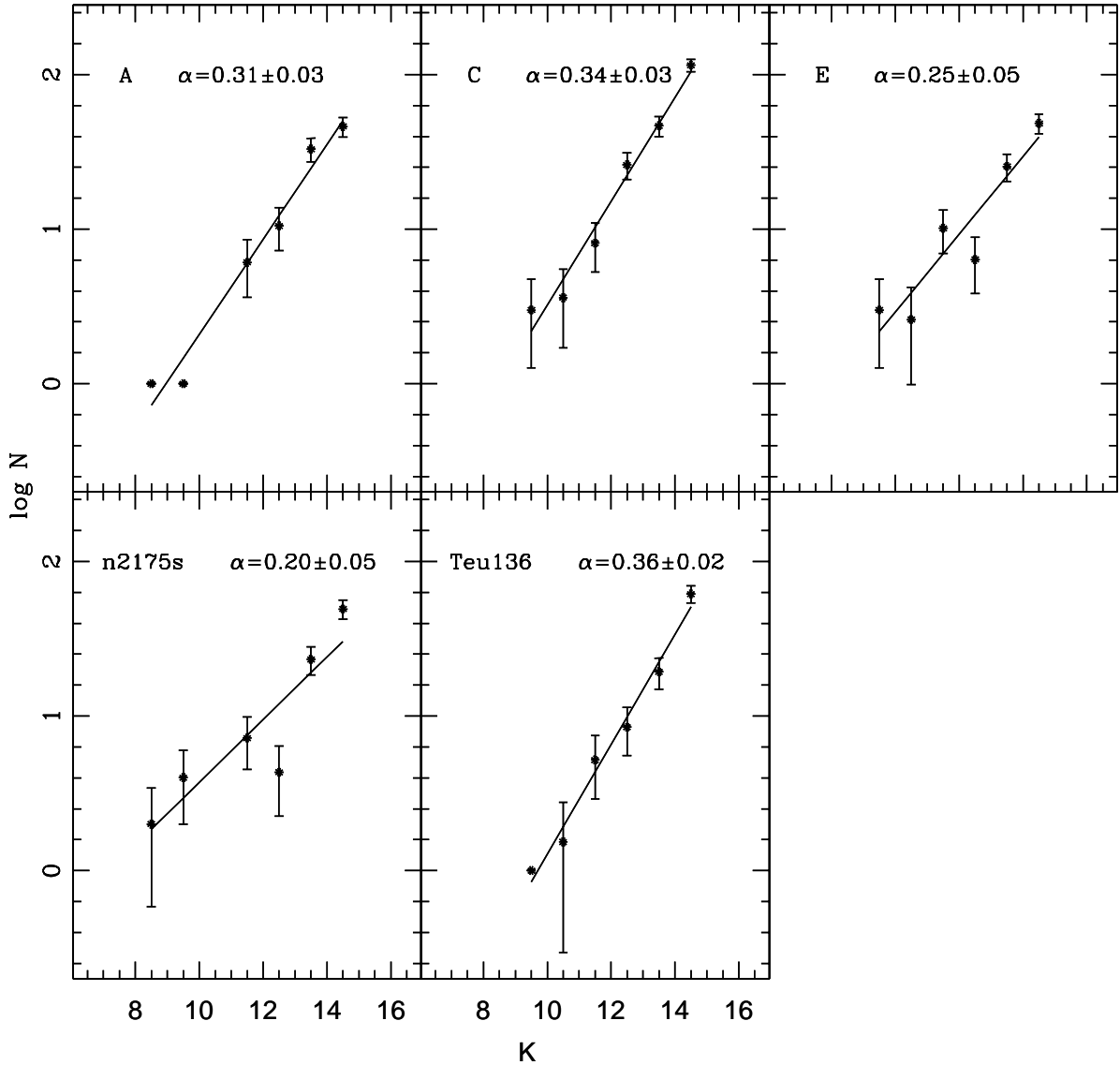}
\caption{KLFs derived  for the sub-regions in Sh2-252 after subtracting the  field star contamination (see the text).
The error bars represent  the $\pm$$\sqrt{N}$ errors. The  linear  fits are represented by the straight  lines and the 
slopes obtained are given  in each figure.}
\label{klf}
\end{figure}


The KLFs of young embedded clusters are known to follow power-law
shapes  (Lada et al. 1991; 1993), which is  expressed as:

\begin{center}
${{ \rm {d} N(K) } \over {\rm{d} K }} \propto 10^{\alpha K}$
\end{center}

where ${ \rm {d} N(K) } \over  {\rm{d} K }$ is the number of stars per 1 mag   
bin  and   $\alpha$  is  the  slope  of   the  power law. We counted the number stars per unit
magnitude interval  after correcting for the field star contribution. We then fit a power law to the data.
The resulting field star subtracted KLFs for the five sub-regions of Sh2-252 are shown in Fig. \ref{klf}. 
 The slopes of the KLFs of regions A, C, E, NGC 2175s  and Teu 136 have been obtained as 0.31$\pm$0.03, 
0.34$\pm$0.03, 0.25$\pm$0.05, 0.20$\pm$0.05 and 0.36$\pm$0.02, respectively. Within  errors,
the KLFs for all the five regions seem to match with each other. 

KLFs of  different ages are known to
have different peak magnitudes and slopes (Muench et al. 2000) and hence the KLF slope could be an 
age indicator of  young clusters. For clusters up to 10 Myr 
old, the KLF slope  gets steeper as the cluster gets older  (Ali \& Depoy 1995;
Lada \& Lada 1995). However, there exists  no precise  age - KLF relationship
in the literature due to  huge uncertainties  in their correlation (Devine et al. 2008).
There are many studies on KLF of young clusters. 
The studies by Blum et al. (2000); Figuer\^{e}do et al. (2002); Leistra et al. (2005; 2006) and Devine at al. (2008)
indicate that the KLF slope  varies from 0.2 - 0.4 for  clusters younger than 5 Myr, which is
in agreement with the KLFs obtained for the sub-regions of Sh2-252. 
The KLFs of the sub-regions  of Sh2-252  are worth comparing with the recent studies of
star forming regions viz; NGC 1893 (Sharma et al. 2007), Be 59 (Pandey et al. 2008), Stock 8 and NGC 1624
(Jose et al. 2008; 2011), since  all the KLFs are obtained by using a similar technique. The slopes
of the KLFs obtained for the sub-regions in Sh2-252 are comparable with those obtained for NGC 1893 
($\alpha = 0.34\pm0.07$), Stock 8 ($\alpha = 0.31\pm0.02$) Be 59  ($\alpha = 0.27\pm0.02$) and 
NGC 1624 ($\alpha = 0.30\pm0.06$).
 
\subsection{Initial mass functions of  sub-regions in Sh2-252}

One of the most fundamental disciplines of astrophysical research is the origin of stars and stellar masses. 
The distribution of stellar masses that form in a star formation event in a given volume of space is called 
IMF and is one of the most important measurable quantities in star formation studies. Together with star 
formation rate, the IMF dictates the evolution and fate of star  clusters and galaxies 
(Kroupa 2002). Young  clusters and star forming regions are important objects to study the  IMF 
since their mass functions (MFs) can be considered  as IMFs  as they  are too young  to loose a significant number of 
the members either  by dynamical  or  stellar evolution.  

The MF is often expressed by the power law,  $N (\log m) \propto
m^{\Gamma}$  and the slope of the MF is given as

$$ \Gamma = d \log N (\log m)/d \log m $$

\noindent where $N  (\log m)$ is the  number of stars per  unit
logarithmic mass interval.  For the mass range $0.4  < M/M_{\odot} \le 
10$,  the classical   value derived  by Salpeter  (1955) is $\Gamma = -1.35$.

One of the main objectives of this work is to measure the MFs of the sub-regions of
Sh2-252. As discussed in section \ref{optcc}, the majority of the sources lying above the 5 Myr
isochrone of the CMDs could be probable PMS members of the region and most of them have a  mass
range of 0.3 - 2.5M$_{\odot}$. We calculated the MFs of these PMS sources falling above 5 Myr and
mass range of 0.3 - 2.5M$_{\odot}$ for the five sub-regions. To remove the contamination due to
field stars from the PMS sample, we statistically subtracted the contribution of field stars
from the observed CMDs of sub-regions (see Fig. \ref{vivall}) using the procedure described in our
earlier studies (e.g., Pandey et al. 2008, Jose et al. 2008; 2011). After statistically subtracting
the field star contribution, we used all the candidate PMS sources lying above the 5 Myr
isochrone and having mass in the range of 0.3 - 2.5M$_{\odot}$ to calculate the MFs. The MFs for
the candidate PMS sources were obtained by counting the number of stars in various mass
bins, shown as evolutionary tracks by Siess et al. (2000) in Fig. \ref{viv}. Necessary corrections
for data incompleteness as a function of magnitude  were taken
into account to calculate the MFs.  The CFs  given in Table \ref{cftable} for regions C and NGC 2175s 
were used to correct for their data incompleteness whereas the average value of the CFs for regions C and NGC 2175s
was applied to regions A, E and Teu 136 to  correct for their data incompleteness. The resulting field star decontaminated and 
completeness corrected MFs of the PMS sources of the five sub-regions of Sh2-252 
region are shown in Fig. \ref{mf}.  The slopes, $\Gamma$,  of the MFs in the mass range
$0.3 \le M/M_{\odot}<2.5$  are found to be -1.33$\pm$0.51, -2.39$\pm$0.15, -1.40$\pm$0.30, -1.65$\pm$0.28,
-1.82$\pm$0.26, respectively, for regions A, C, E, NGC 2175s and Teu 136. The slopes
of the MFs of the regions A and E are in good agreement with that of the
Salpeter (1955) value, and in view of large errors, those in
the case of NGC 2175s and Teu 136 can also be considered comparable to the
Salpeter value. However, in the case of the region C, the MF seems to be
steeper than the Salpeter value. In order to check this we also calculated the MF 
using $J$-band luminosity function (JLF). The procedure for this calculation has been mentioned in
our earlier paper (Jose et al. 2011). The MF thus obtained for region C from JLF 
is shown in Fig. \ref{mf} as dashed line and the slope obtained is -2.10$\pm$0.54. Considering the large
uncertainty in the IMF calculation, we conclude that all the regions have IMF slopes, which do not
significantly deviate from the salpeter value.

The shape of the stellar IMF and whether it is universal or not are key issues in astrophysics. 
For clusters within 2 kpc, there is no compelling evidence for variations in the
stellar IMF with respect to their environment (e.g. Meyer et al. 2000; Kroupa 2002; Chabrier
2005). We have been  pursuing studies of various star forming regions in the galaxy, hence a comparative study of their IMFs
obtained using similar techniques will give useful information
on IMFs. Our recent analysis of young clusters (age  2-4 Myr), viz., NGC
1893 (Sharma et al. 2007), Stock 8, NGC 1624 (Jose et al. 2008, 2011) and W5
E (Chauhan et al. 2011) have yielded  values  similar to the Salpeter
value.  In the present study, the MFs of all the sub-regions of Sh2-252
are found to be in general comparable to the Salpeter value.

\begin{figure}
\centering
\includegraphics[scale = .8,  trim = 0 0 0 0, clip]{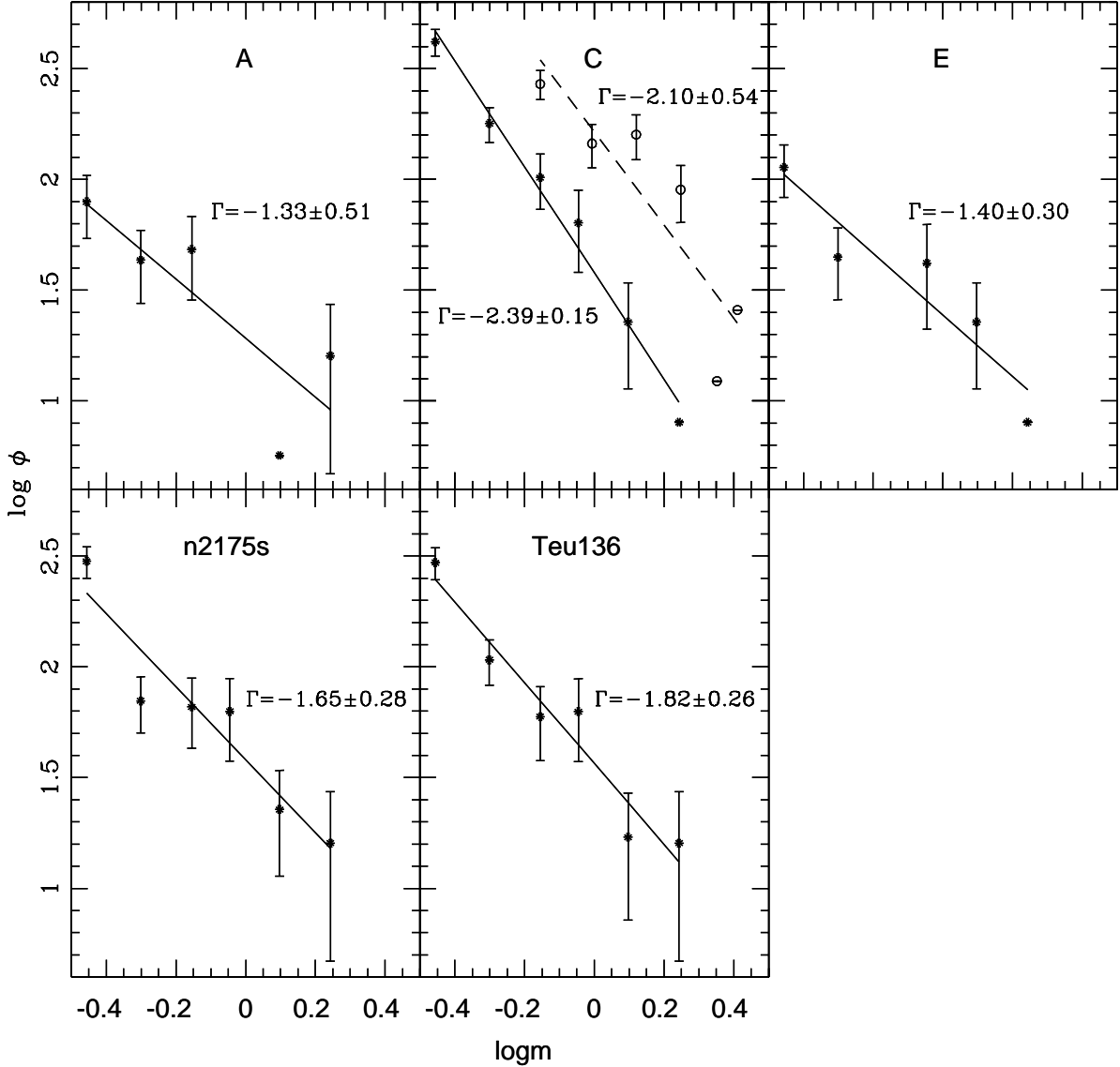}

\caption{MFs obtained from the optical data for the sub-regions of Sh2-252 after correcting for the field star contamination 
and data incompleteness. The $\phi$ represents $N$/dlog $m$ and the error bars represent  $\pm$$\sqrt{N}$ errors. 
The continuous lines show least-squares fits  to the mass ranges described in the text.  The 
value of the slopes obtained are given in each figure.  The MF obtained from JLF for region C is shown in the second panel 
and the least-squares fit is shown by using the dashed line. }
\label{mf}
\end{figure}

\section{Summary}

In this paper we studied the stellar contents of the \hii region Sh2-252 which itself contains four C\hii regions
such as A, B, C and E and two clusters NGC 2175s and Teu 136. We used the  deep optical 
photometry in $UBVRI$ bands, slit and slitless spectroscopic observations  along with 
the  $JHK$ data from 2MASS for our analyses. An attempt has been made to identify and classify the massive members,
to determine the fundamental parameters of Sh2-252 as well as to obtain the age and mass distribution  of candidate PMS 
sources and finally to constrain  the KLFs and IMFs of the sub-regions of Sh2-252.

We have carried out optical spectroscopy  of fifteen bright sources  of the region, out of which,
eight  have been identified as massive members of spectral type earlier than B3. The C\hii regions A, B, C and E
are found to have at least one star of spectral type  earlier than B3, whereas, the small optically bright
cluster NGC 2175s is found to have four  stars of  spectral type earlier than B3 around  its center. We have also identified the
probable candidate ionizing sources of the  C\hii regions A, B, C and E. From the spectro-photometric 
analysis, we derived the average distance of the region as 2.4 $\pm$ 0.2 kpc. The distance  estimated from the four bright stars
of NGC 2175s is found to be  in good agreement with  the average distance of Sh2-252, which supports the association of NGC 2175s
with the Sh2-252 complex. In this paper, we also report that the NIR embedded cluster Teu 136, located at the eastern edge of the complex 
as a sub-cluster of  Sh2-252. Based on the optical colour-colour diagram and spectroscopic properties,  the reddening $E(B-V)$ 
of the massive members of the region  is found to  vary between 0.35 mag to 2.1 mag.

Using slitless spectroscopy survey, we have identified 61 $H\alpha$ emission sources in the region. We obtained optical photometry 
for 211 candidate YSOs of the region. The distribution of these candidate
YSOs on the $V/(V-I)$ CMD shows that a majority of them  have an age distribution between 0.1 - 5 Myr and masses in the range of 
0.3 - 2.5 M$_{\odot}$. The stellar surface density distribution in $K$-band shows that there is at least five sub-clusters, each associated with
the regions A, C, E, NGC 2175s and Teu 136.  The $(J-H)/(H-K)$ C-C diagrams of the individual regions show that the regions A, C and E 
have more reddened population when compared to the clusters NGC 2175s and Teu 136, which are located towards the east of Sh2-252. 
The optical CMDs of the individual regions also show that the candidate PMS sources are distributed between 0.1 - 5 Myr, 
suggesting an age spread  for them. 

 We also calculated  the KLFs of the sub-regions  A, C, E, NGC 2175s  and Teu 136 and have estimated  as
0.31$\pm$0.03, 0.34$\pm$0.03, 0.25$\pm$0.05, 0.20$\pm$0.05 and 0.36$\pm$0.02, respectively.
Within errors, the KLFs of all the sub-regions are similar and comparable to that of young clusters 
of age $<$ 5 Myr.  We also derived the MFs of PMS sample of the individual regions in  the mass range 
of $0.3 - 2.5 M_{\odot}$.  The slopes of the MFs of all the sub-regions are found to match  with the Salpeter value.

\section{Acknowledgments}
Authors are thankful to the anonymous referee for the  useful comments
which has improved contents and presentation of the paper significantly. 
We thank the staff of the Kiso Observatory, Japan, ARIES, Naini Tal, India  and IAO, Hanle and its remote control station
at CREST, Hosakote for their assistance during the observations. This
publication also makes use of data from the Two Micron All Sky Survey, which is a joint project of
the University of Massachusetts and the Infrared Processing and Analysis Center/California
Institute of Technology, funded by the National Aeronautics and Space Administration and
the National Science Foundation. JJ and NC are thankful for the financial support for this study
through a stipend from CSIR and DST, India. AKP is thankful to DST (India) and JSPS (Japan) for 
providing funds to visit KISO observatory to carry out the observations. 

\section*{REFERENCES}
Ali B.,  Depoy D. L., 1995, AJ, 109, 709\\
Bessell M.,  Brett J. M., 1988, PASP, 100, 1134\\
Blum R. D., Conti P. S.,  Damineli A., 2000, AJ, 119, 1860\\
Bonatto C., Bica E., 2011, MNRAS, 414, 3769\\
Carpenter J. M., 2001, AJ, 121, 2851\\
Carraro G., Vazquez R. A., Moitinho A.,  Baume G., 2005, ApJ, 630, L153\\
Chabrier G., 2005, The Initial Mass Function 50 Years Later, 327, 41\\
Chauhan N., Pandey A. K., Ogura K., Jose J., Ojha D. K., Samal M. R.,  Mito H., 2011, MNRAS, 415, 1202\\
Conti P.S.,  Alschuler W. R., 1971, ApJ, 170, 325\\
Cohen J. G., Frogel J. A., Persson S. E.,  Ellias J. H., 1981, ApJ, 249, 481\\
Cutri R. M., Skrutskie M. F., van Dyk S. et al., 2003, The IRSA 2MASS All Sky Point Source Catalog, NASA/IPAC 
Infrared Science Archive, http://irsa.ipac.caltech.edu/applications/Gator/\\
Chavarr\'{i}a-K. C., de Lara E., Hasse I., 1987, A\&A, 171, 216\\
Chavarr\'{i}a-K. C. et al., 1989, A\&A, 215, 51\\
Chavarria L., Mardones D., Garay G., Escala A., Bronfman L.,  Lizano S., 2010, ApJ, 710, 583\\
Churchwell E., 1974, in proceedings of the Second European Regional Meeting in Astron. Mem. Soc. Astron. Ital. 45, 259\\
Clarke C. J., 2007, MNRAS, 376, 1350\\
Dahm S., 2005, AJ, 130, 1805\\
Devine K. E., Churchwell E. B., Indebetouw R., Watson C.,  Crawford S. M., 2008, AJ, 135, 2095\\
Elmegreen B.G.,  Lada C.J., 1977, ApJ, 214, 725\\
Felli M., Habing H. J.,  Isra\"{e}l F.P., 1977, A\&A, 59, 43\\
Figuer\^{e}do E., Blum R. D., Damineli A.,  Conti P. S., 2002, AJ, 124, 2739\\
Garnier R.,  Lortet-Zuckermann M. C., 1971, A\&A, 31, 41\\
Georgelin Y. P.,  Georgelin Y. M., 1970, A\&A, 6, 349\\
Girardi L., Bertelli G., Bressan A., Chiosi C., Groenewegen M. A. T. et al., 2002, A\&A, 391, 195\\
Grasdalen G. L., Carrasco L., 1975, A\&A, 43, 259\\
Gutermuth R. A., Megeath S. T., Myers P. C., Allen L. E., Pipher J. L.,   Fazio G. G., 2009, ApJS, 184, 18\\
Haikala L. K., 1994, A\&A, 108, 643\\
Hernandez J., Calvet N., Hartmann L., Briceno C., Sicilia-Aguilar A., Berlind P., 2005, AJ, 129, 856\\
Haisch K. E., Lada E. A., Lada C. J., 2000, AJ, 120, 1396\\
Haisch K. E., Lada E. A.,  Lada C. J., 2001, AJ, 121, 2065\\
Hillenbrand L. A., Strom S. E., Vrba F. J.,  Keene J., 1992, ApJ, 397, 613\\
Hunter T. R., Testi L., Taylor G. B., Tofani G., Felli M.,  Phillips T. G., 1995, A\&A, 302, 249\\
Jacoby G. H., Hunter D. A.,  Christian C. A., 1984, ApJS, 56, 257\\
Jose, J. et al., 2008, MNRAS, 384, 1675\\
Jose, J. et al., 2011, MNRAS, 411, 2530\\
K\"{o}mpe C., Joncas G., Baudry A.,  Wouterloot J. G. A., 1989, A\&A, 221, 295\\
Koposov S.E., Glushkova E.V.,  Zolotukhin I.Yu., 2008,  A\&A, 486, 771\\
Koornneef J., 1983, A\&A, 128, 84\\
Kroupa P., 2002, SCIENCE, 295, 82\\
Lada C. J.,  Wooden D., 1979, ApJ, 232, 158\\
Lada C. J., Lada E. A., 1991, in ASP Conf. Ser. 13, The Formation and Evolution of Star Clusters, ed. K. Janes (San Francisco: ASP), 3 \\
Lada C. J.,  Adams F. C., 1992, ApJ, 393, 278\\
Lada C. J., Young T., Greene T., 1993, ApJ, 408, 471\\
Lada E . A .,  Lada C . J., 1995, AJ , 109, 1682\\
Lada C. J.,  Lada E. A., 2003, ARA\&A, 41, 57\\
Landolt A.U., 1992, AJ, 104, 340\\
Lee H. T., Chen W. P., Zhang Z. W., Hu J.Y., 2005, ApJ, 624,808\\
Leistra A., Cotera A. S., Leibert J.,  Burton M., 2005, AJ, 130, 1719\\
Leistra A., Cotera A. S.,  Liebert J., 2006, AJ, 131, 2571\\
Mart\'{i}n-Hern\'{a}ndez N. L., van der Hulst J. M.,  Tielens A. G. G. M., 2003,  A\&A, 407, 957\\
Meyer M., Calvet N.,  Hillenbrand L. A., 1997, AJ, 114, 288\\
Meyer M. R., Adams F. C., Hillenbrand L. A., Carpenter J. M., Larson R. B., 2000, Protostars and Planets IV, 121\\
Meynet G., Maeder A., Schaller G., Schaerer D.,  Charbonnel C., 1994, A\&AS, 103, 97\\
Muench A. A., Lada E.A., Lada C.J., 2000, ApJ, 553, 338\\
Neckel T., Klare G., Sarcander M., 1980, Bull. Inform. Centre, des Donees Stellaires, 19\\
Ojha D. K., Tamura M., Nakijama Y. et al., 2004a, ApJ, 608, 797\\
Ojha D. K., Tamura M., Nakajima Y. et al., 2004b, ApJ, 616, 1042\\
Panagia, N., 1973, AJ, 78, 929\\
Pandey A.K., Nilakshi, Ogura K., Sagar R.,  Tarusawa K., 2001, A\&A, 374, 504\\
Pandey A. K., Sharma S.,  Ogura K., 2006, MNRAS, 373, 255\\
Pandey A. K., Sharma S., Ogura K., Ojha D. K., Chen W. P. et al., 2008, MNRAS, 383, 1241\\
Pismis P., 1970, Boletin Observatorio Tonantzintla Tacubaya, 5, 219\\
Pismis P., 1977, Rev. Mex. Astron. Astrofis., 2, 59\\
Reed B. C., 2003, AJ, 125, 2531\\
Reid M. J., Menten K. M., Brunthaler A., Zheng X. W., Moscadelli L.,  Xu Y., 2009, ApJ, 693, 397\\
Robin A. C., Reyle C., Derriere S.,  Picaud S., 2003, A\&A, 409, 523\\
Robitaille T. P., Whitney B. A., Indebetouw R., Wood K.,   Denzmore P., 2006, ApJS, 167, 256\\
Salpeter E. E., 1955, ApJ, 121, 161\\
Sanchawala K., Chen W. P., Ojha D. et al., 2007, ApJ, 667, 963\\
Schmidt-Kaler Th. 1982, in Landolt-Bornstein, Vol. 2b, ed. K. Schaifers, H. H. Voigt, \& H. Landolt (Berlin: Springer), 19\\
Sharpless S., 1959, ApJS, 4, 257\\
Sharma S., Pandey A. K., Ojha D. K., Chen W. P., Ghosh S. K., Bhatt B. C., Maheswar G.,  Sagar R., 2007, MNRAS, 380, 1141\\
Siess L., Dufour E.,  Forestini M., 2000, A\&A, 358, 593\\
Sugitani K. et al., 2002, ApJ, 565, L25\\
Szymczak M., Hrynek G.,  Kus A. J., 2000, A\&AS, 143, 269\\
Tej A., ojha D. K., Ghosh S. K., Kulkarni V. K., Verma R. P., Vig S.,  Prabhu T. P., 2006, A\&A, 452, 203\\
Torres-Dodgen Ana V., Weaver W. B., 1993, PASP, 105, 693\\
Walborn N.R., 1972, AJ, 77, 312\\
Walborn N. R., Fitzpatrick E. L., 1990, PASP, 102, 379\\
Zinnecker H., McCaughrean M. J.,  Wilking B. A., 1993, in Protostars and Planets III, ed. E. Levy \& J. Lunine (Tucson: Univ. Arizona Press), 429\\

\bsp

\end{document}